# Advances in Cryogenic Avalanche Detectors

## A. Buzulutskov

*Budker Institute of Nuclear Physics SB RAS, Lavrentiev avenue 11, 630090 Novosibirsk, Russia*
*Novosibirsk State University, Pirogov street 2, 630090 Novosibirsk, Russia*
*E-mail:* `A.F.Buzulutskov@inp.nsk.su`

ABSTRACT: Cryogenic Avalanche Detectors (CRADs) are referred to as a new class of noble-gas detectors operated at cryogenic temperatures with electron avalanching performed directly in the detection medium, the latter being in gaseous, liquid or two-phase (liquid-gas) state. Electron avalanching is provided by Micro-Pattern Gas Detector (MPGD) multipliers, in particular GEMs and THGEMs, operated at cryogenic temperatures in dense noble gases. The final goal for this kind of detectors is the development of large-volume detectors of ultimate sensitivity for rare-event experiments and medical applications, such as coherent neutrino-nucleus scattering, direct dark matter search, astrophysical (solar and supernova) neutrino detection experiments and Positron Emission Tomography technique. This review is the first attempt to summarize the results on CRAD performances obtained by different groups. A brief overview of the available CRAD concepts is also given and the most remarkable CRAD physics effects are discussed.

**Contents**

## 1. Introduction

Over the past decade there has been a growing interest in so-called "Cryogenic Avalanche Detectors". This term has been known since 2003 [1]; in the wide sense it defines a new class of noble-gas detectors operated at cryogenic temperatures with electron avalanching performed directly in the detection medium. The detection medium can be in a gaseous, liquid or two-phase (e.g. liquid-gas) state. The ultimate goal for this kind of detectors is the development of large-volume detectors of ultimate sensitivity, i.e. operated in single-electron counting mode at extremely low noise, for rare-event experiments and other (e.g. medical imaging) applications. These include coherent neutrino-nucleus scattering, direct dark matter search, astrophysical (solar and supernova) neutrino detection experiments and Positron Emission Tomography (PET) technique. A typical deposited energy in these experiments might be rather low: of the order of 0.1, 1-10, ≥100 and 500 keV, respectively. Accordingly, the primary ionization and/or scintillation signals should be amplified in dense noble-gas media at cryogenic temperatures.

Earlier attempts to obtain high and stable electron avalanching directly in noble gases and liquids at cryogenic temperatures, using "open-geometry" gaseous multipliers, have not been very successful: rather low gains ($\leq$10) were observed in liquid Xe [2],[3],[4] and Ar [2],[5] and low gains ($\leq$100) in gaseous Ar [6] and He [7] at low temperatures, using wire, needle or micro-strip proportional counters. Moreover, two-phase detectors with wire chamber readout, which initially seemed to solve the problem, turned out to have unstable operation in the avalanche mode due to vapour condensation on wire electrodes [8].

The problem of electron avalanching in cryogenic noble-gas detectors has been solved [1] after introduction of Micro-Pattern Gas Detectors (MPGDs), namely those of hole-type: Gas Electron Multipliers (GEMs) [9] and thick GEMs (THGEMs) [10]. Contrary to wire chambers and other "open geometry" gaseous multipliers, cascaded GEM and THGEM structures have a unique ability to operate in dense noble gases at high gains [11], including at cryogenic temperatures and in the two-phase mode [12].

Consequently at present, the basic idea of Cryogenic Avalanche Detectors in the narrow sense is that of the combination of MPGDs with cryogenic noble-gas detectors, operated in a gaseous, liquid or two-phase mode. We call such detectors "CRyogenic Avalanche Detectors" and suggest the following short-name for those - CRADs; it will be used throughout in the following. This review is the first attempt to summarize the results on CRAD performances obtained by different groups, presenting those in a systematic way. A brief overview of the available CRAD concepts is also given and the most remarkable CRAD physics effects are discussed.

## 2. Cryogenic Avalanche Detector (CRAD) concepts: brief overview

There are at least a dozen of different CRAD types developed by different groups over the past 8 years. In this chapter this variety is systemized: a brief overview of CRAD concepts is given, namely that of the basic CRAD concepts and CRAD concepts related to some experimental projects. The reasons and motivations to develop either one or another concept will be clarified in section 2.4 when discussing these CRAD-related projects.

Each CRAD concept contains MPGD multiplier as a basic element; in most cases it is a GEM or THGEM multiplier. GEM is a thin insulating film, metal clad on both sides, perforated by a matrix of micro-holes, in which gas amplification occurs under the voltage applied across the film [9]. THGEM is a similar, though more robust structure with ten-fold expanded dimensions [10].

Another basic element for a large number of concepts is a two-phase electron-emission detector. In conventional two-phase detectors [13],[14],[15],[16],[17], the electrons from the primary ionization drift in the liquid to its surface; under moderate electric field they are emitted into the gas phase, where they produce secondary scintillations (electroluminescence) in the vacuum ultraviolet (VUV) spectral range. In conventional two-phase detectors the primary and secondary scintillation signals are recorded in the detector volume using PMTs operating at cryogenic temperatures.

### 2.1 Original CRAD concept

The original CRAD concept was first mentioned in 2002 [18] and then finally introduced in 2003 [1], at Budker INP (see Fig. 1): electron avalanching at cryogenic temperatures is performed in pure noble gases using a hole-type multiplier, namely that of GEM, either in a gaseous or two-phase mode. In the latter case the conventional two-phase electron-emission detector is provided with electron avalanching in the gas phase: the primary ionization electrons produced in noble liquid are emitted into the gas phase by an electric field, where they are multiplied in saturated vapour above the liquid using a cascaded GEM multiplier. The proof of principle of this concept was demonstrated in 2003-2004 in two-phase Kr and in gaseous He, Ar and Kr at temperatures down to 120 K [1],[19]. In 2005-2006 this concept was proved in gaseous He and Ne at temperatures down to 2.6 K [20],[21] and in two-phase Ar and Xe [22].

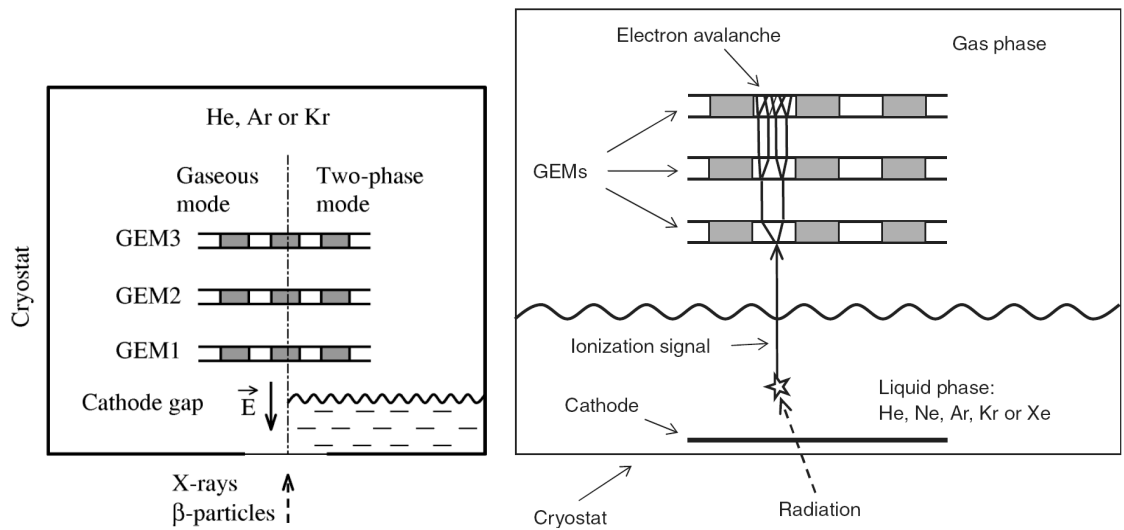

Fig. 1. Original Cryogenic Avalanche Detector (CRAD) concept introduced in 2002-2003 [1],[18]: gaseous (left) and two-phase (left and right) CRADs with GEM multiplier charge readout. The figures are taken from [1] and [22].

Later on, the original CRAD concept was elaborated: it was suggested to provide CRADs with new features. Those of the most significance are listed below in order of introduction, with appropriate references relevant to the concept introduction and its proof of principle:

- THGEM (or LEM) multiplier charge readout, in two-phase CRADs [23],[24],[25],[26],[27],[28];
- MPGD-based cryogenic Gaseous Photomultiplier (GPM) separated by window from the noble liquid, in liquid CRADs [29],[30],[31];
- primary scintillation signal readout using CsI photocathode on the first GEM, in two-phase CRADs [22],[32];
- CCD optical readout of the GEM multiplier, in gaseous and two-phase CRADs [33],[34];
- Geiger-mode APD (GAPD or SiPM) optical readout of the THGEM multiplier, in two-phase, liquid and gaseous CRADs [26],[28],[35],[36],[37],[38].

The gallery of CRAD concepts based on these and other features, developed since 2003, is shown in Fig. 1a; these will be detailed in the following sections. But at first a condition "sine qua non" crucial for most CRAD concepts will be discussed, namely that of the high gain operation of MPGD multipliers in dense noble gases.

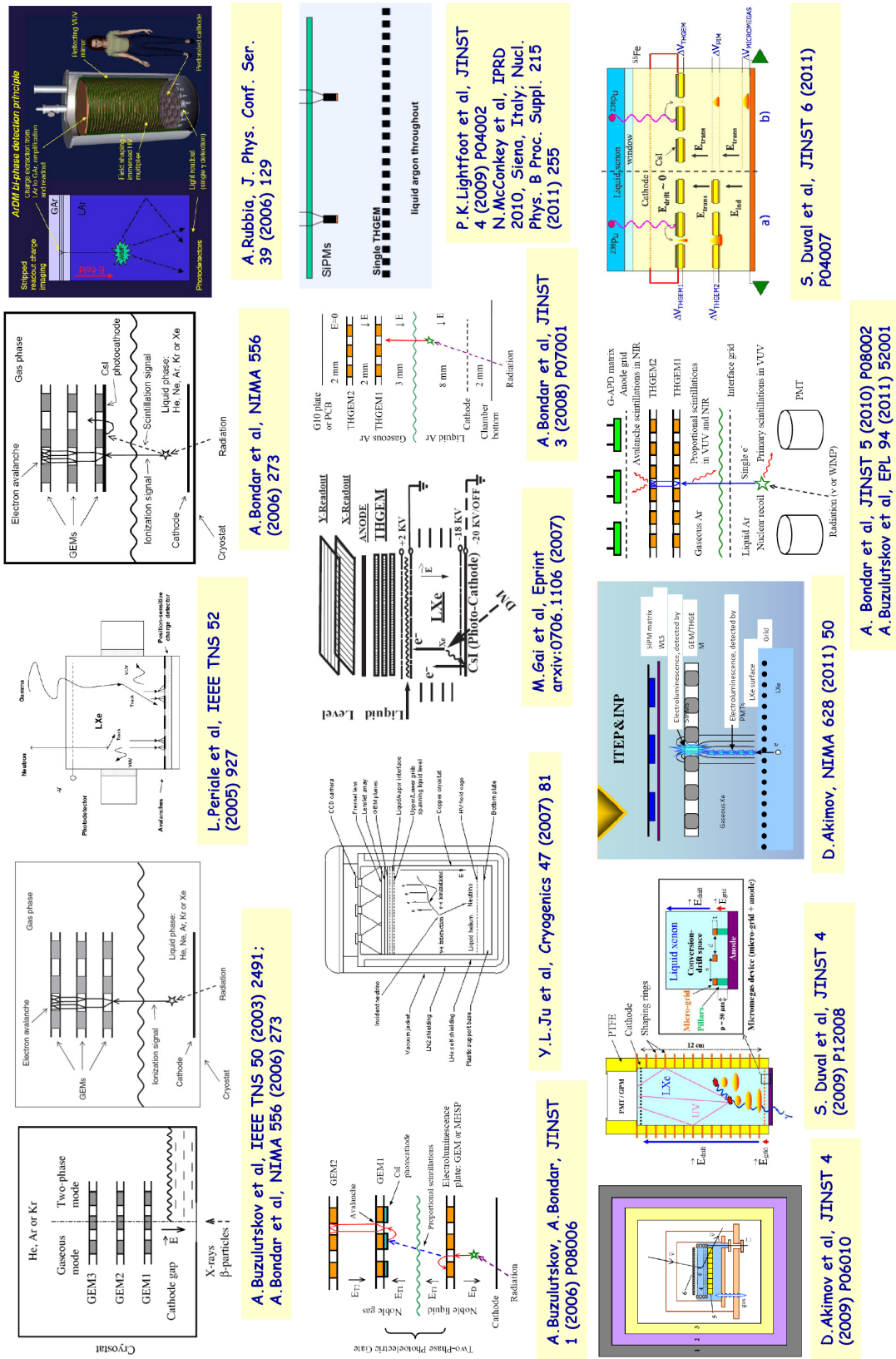

Fig. 1a. CRAD concept gallery. The concepts are shown in order of introduction in 2003-2011.

## 2.2 Condition "sine qua non": MPGD multiplier operation in dense noble gases

In most CRAD concepts, the MPGD multiplier should be able to operate at high gains in dense pure noble gases, in particular in saturated vapour and at cryogenic temperatures, to provide high detection efficiency for primary ionization and scintillation signals. This was a real challenge for gaseous multipliers, since it was generally believed that high gains were inaccessible in pure noble gases due to considerably enhanced secondary avalanche effects, in particular in the absence of molecular quenching additives.

Fortunately, unlike open-geometry gaseous multipliers (e.g. wire chambers), hole-type MPGD multipliers permit attaining high charge gains in "pure" noble gases, presumably due to considerably reduced photon-feedback effects. This remarkable property was discovered in 1999-2000 jointly by Budker INP, Weizmann Institute and CERN groups [11],[39]. It was demonstrated for cascaded GEM multipliers operated with high gains ($\geq 10^4$) first in Ar and its mixtures with other noble gases [11],[39], then in all other noble gases at normal and high pressure [18],[40],[41]: see Figs. 2 and 3 showing the appropriate gain characteristics and maximum gains. This unique property was interpreted introducing the concept of the "avalanche confinement" within the GEM hole [11],[42],[43]. In Fig. 3, the striking difference in maximum gain dependence on pressure between heavy (Ar, Kr, Xe) and light (He, Ne) noble gases is explained by the complementary avalanche mechanism available in He and Ne, namely by that of the Penning effect in uncontrolled impurities; this will be discussed in detail in sections 3.1 and 4.4.

Later on, the high gain operation in dense noble gases was demonstrated in other MPGD multipliers: in THGEMs and resistive THGEMs (RETHGEMs) in Ar, Kr and Xe [44],[45],[46],[47],[48], in Micro-Hole and Strip Plates (MHSPs) in Ar, Kr and Xe [49] and in Micromegases (MMs) in Xe [50]. Their gain characteristics are shown in Figs. 4 and 5. One can see that compared to GEMs in terms of the maximum gain, THGEMs are similar to GEMs, MHSPs benefit in Kr and Xe and at higher pressures, MMs lose in Xe at 1 atm but benefit at higher pressures (>3 atm).

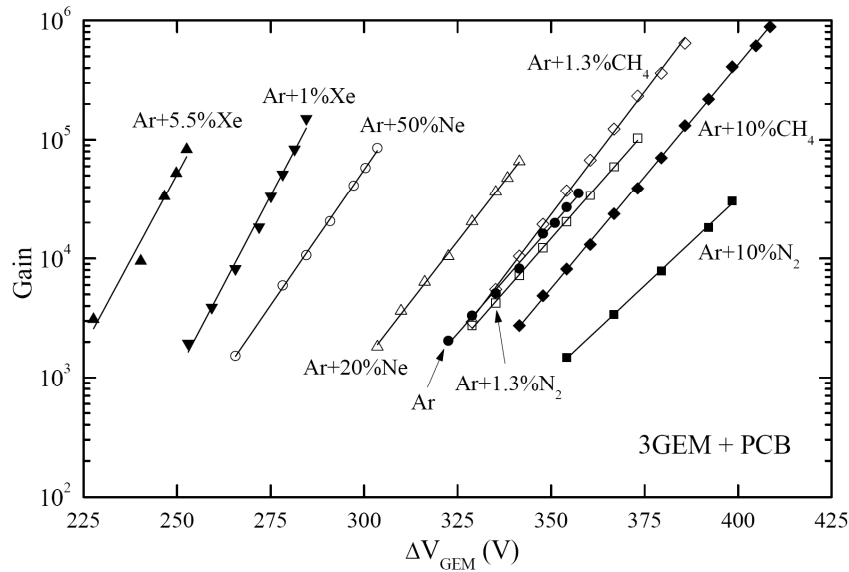

Fig. 2. Gain characteristics of a triple-GEM multiplier in Ar and its mixtures with Ne, Xe, $N_2$ and $CH_4$ at room temperature and atmospheric pressure [11].

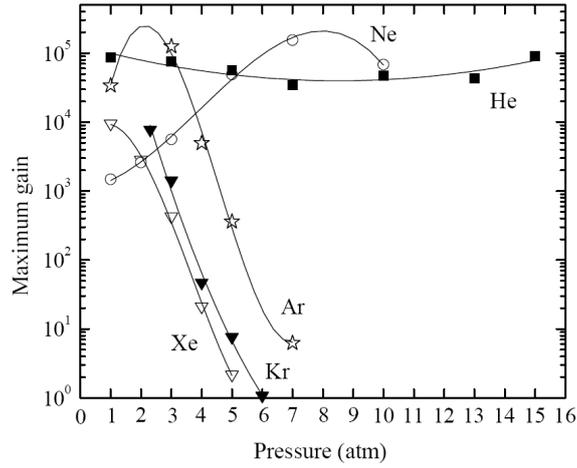

Fig. 3. Maximum gains of triple-GEM multipliers as a function of pressure in He, Ne, Ar, Kr and Xe at room temperature [18].

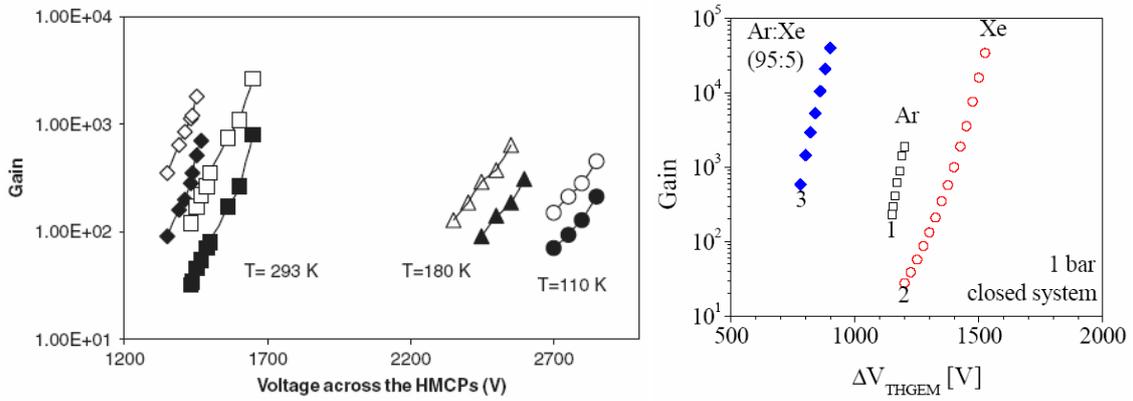

Fig. 4. Gain characteristics of THGEM multipliers at 1 atm. Left: for single- (solid symbols) and double- (open symbols) THGEM in Xe (squares and triangles) or Ar (rhombus and circles) [44]. Right: for double-THGEM in Ar and Xe at room temperature [46].

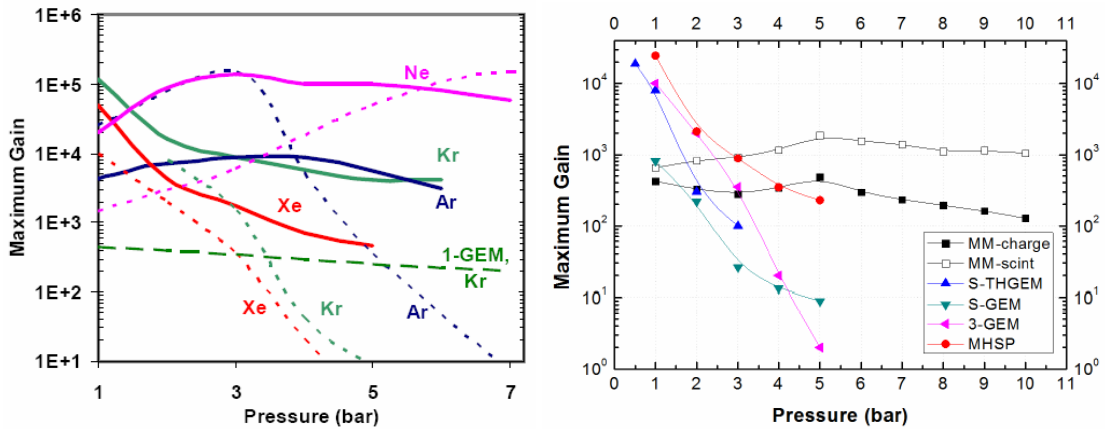

Fig. 5. Maximum MPGD gain as a function of pressure at room temperature. Left, solid curves: for MHSP in Ne, Ar, Kr and Xe [49]. Right: for MM in Xe [50].

## 2.3 Basic CRAD concepts

In this section, a brief overview of basic CRAD concepts is given.

**Gaseous CRAD with hole-type multiplier charge readout.**

The concept was introduced in 2003 [1] (see Fig. 1, left): the primary ionization electrons produced in dense pure noble gas at cryogenic temperatures, are multiplied in the same medium using a cascaded hole-type multiplier, in particular that of GEM and THGEM.

The proof of principle of this concept was demonstrated:

- first for GEM multipliers in He, Ar and Kr at temperatures down to 120 K [1],[19];
- then for GEM multipliers in He and Ne at temperatures down to 2.6 K [20],[21];
- then for THGEM and RETHGEM multipliers in Ar and Xe at temperatures down to 100 K [28],[44],[45].

**Two-phase CRAD with GEM-multiplier readout.**

For the first time the concept was mentioned in 2002 as motivation for the study of GEMs in high-pressure noble gases [18]. The concept was finally introduced in 2003 [1] (see Fig. 1, left and right): the primary ionization electrons produced in noble liquid, are emitted into the gas phase by an electric field where they are multiplied in saturated vapour above the liquid using a cascaded GEM multiplier. In 2006 this concept was elaborated [22],[32] (see Fig. 6): the two-phase CRAD was supplied with the GEM multiplier, reading out both the charge (ionization) signal and that of primary scintillations; for the latter a CsI photocathode on the first GEM was used, thus employing the concept of the windowless cryogenic Gaseous Photomultiplier (GPM).

The proof of principle of this concept was demonstrated:

- first in Kr [1],[19];
- then in Ar and Xe [22] and again in Xe [51];
- then in Ar in single electron counting mode with external trigger [52];
- then in Ar with primary scintillation signal readout using CsI photocathode on the first GEM [32];
- then in Ar demonstrating long-term (1 day) stability [53].

**Two-phase CRAD with THGEM-multiplier charge readout.**

The concept was introduced in 2004 [23] and elaborated in 2008 [24],[25],[27] (see Fig. 7): the primary ionization electrons produced in noble liquid, are emitted into the gas phase by an electric field where they are multiplied in saturated vapour above the liquid using a cascaded THGEM multiplier.

The proof of principle of this concept was demonstrated:

- first in Ar with 2.5×2.5 cm$^2$ THGEM active area (Fig. 7, left) [24];
- then in Ar with 10×10 cm$^2$ THGEM active area and 2D readout tracking (Fig. 7, right) [25],[27];
- then in Ar with 4×4 cm$^2$ THGEM active area [26];
- then in Xe with 2.5×2.5 cm$^2$ THGEM active area [28].

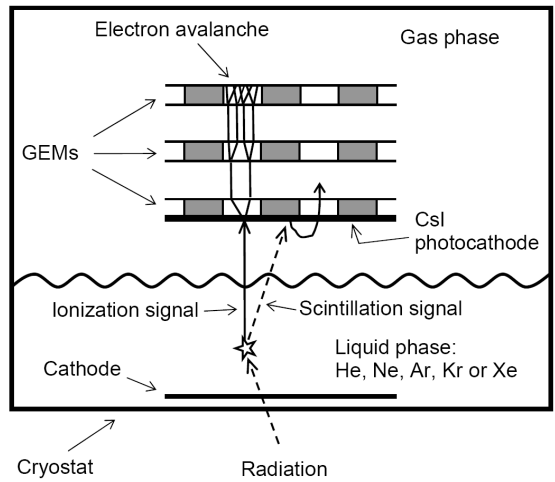

Fig. 6. Two-phase CRAD with GEM multiplier readout of both the charge (ionization) signal and that of primary scintillations [22],[32]; for the latter a CsI photocathode on the first GEM is used (i.e. employing the "windowless cryogenic GPM" concept).

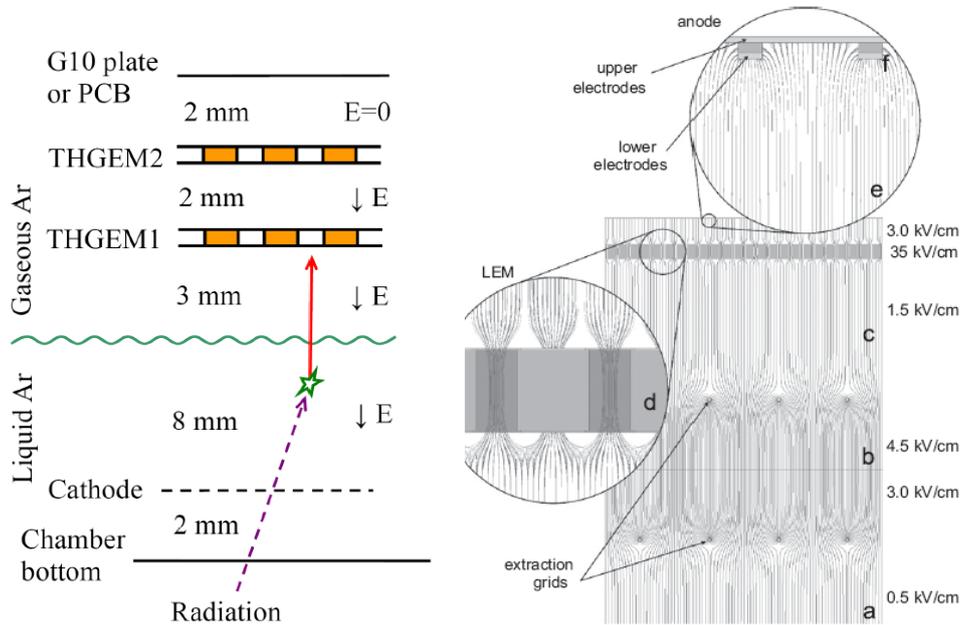

Fig. 7. Two-phase Ar CRADs with THGEM multiplier charge readout, having 2.5×2.5 cm$^2$ active area (left, [24]) and 10×10 cm$^2$ active area and 2D readout (right, [25],[27]).

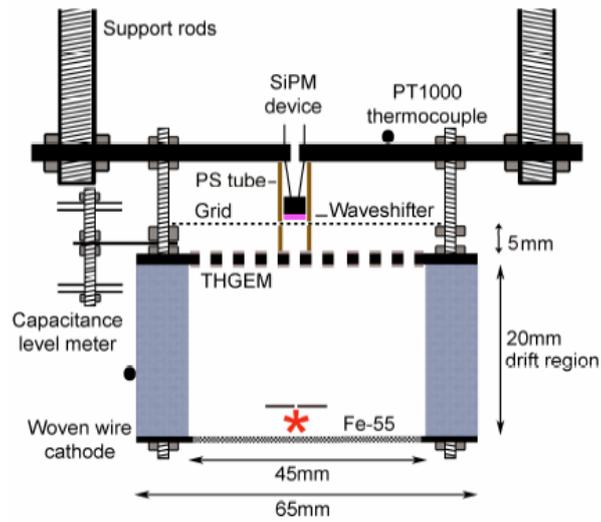

Fig. 8. Two-phase and liquid Ar CRAD with THGEM/GAPD optical readout in the VUV using WLS [26].

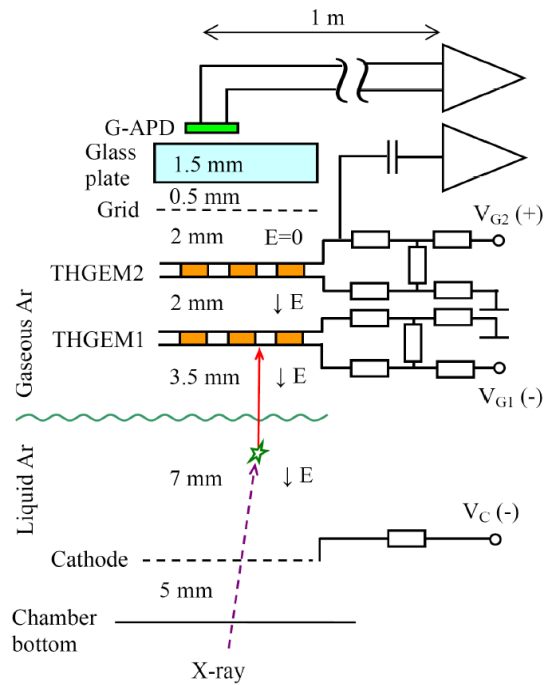

Fig. 9. Two-phase Ar CRAD with THGEM/GAPD optical readout in the NIR [35].

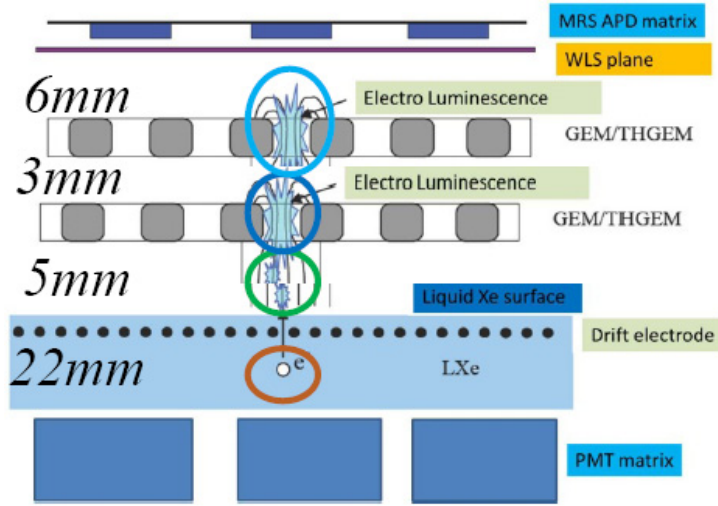

Fig. 10. Two-phase Xe CRAD with THGEM/GAPD-matrix optical readout in the VUV using WLS [38].

**CRAD with optical readout using combined THGEM/GAPD multiplier.**

The concept was introduced in 2009 [26] and elaborated in 2010-2011 [35],[36],[37] (see Figs. 8, 9 and 10): the primary ionization electrons produced in dense noble gas either in a two-phase or gaseous mode, are multiplied using a cascaded THGEM multiplier; then the avalanche scintillations from the THGEM holes are optically read out using GAPDs in the Vacuum Ultraviolet (VUV) or Near Infrared (NIR) region, thus employing the idea of combined THGEM/GAPD multiplier optical readout. In case of readout in the VUV, a Wavelength Shifter (WLS) film is used in front of the GAPD to convert the avalanche scintillation light to the GAPD sensitivity region. In case of readout in the NIR, uncoated GAPDs are used due to their high sensitivity in the NIR.

The proof of principle of this concept was demonstrated:

- first in two-phase Ar with THGEM/GAPD optical readout in the VUV using WLS (Fig. 8) [26];

- then in two-phase Ar with THGEM/GAPD optical readout in the NIR (Fig. 9) [35];

- then in gaseous Xe with THGEM/GAPD optical readout in the NIR [28];

- then in two-phase Xe with THGEM/GAPD-matrix optical readout in the VUV using WLS (Fig. 10) [38].

**Cryogenic Gaseous Photomultiplier (GPM) with CsI photocathode.**

The concept was introduced in 2004-2005 [29],[54]: GPMs with CsI photocathode based on hole-type MPGDs, namely on THGEMs, GEMs or glass GEMs called Capillary Plates (CPs), are operated at cryogenic temperatures in pure noble gases and in their mixtures with quenching additives, either in a sealed mode, i.e. using a window separating GPM from the detection medium, or in a windowless mode, i.e. with GPM operated directly in the detection medium.

The proof of principle of this concept was demonstrated:

- first for CP multipliers in Ar- and He-based mixtures with molecular additives at temperatures down to 80 K [29], in a sealed mode;

- then for THGEM and RETHGEM multipliers in pure Ar and Xe at temperatures down to 100 K, in a windowless mode [44],[45];
- then for triple-GEM multiplier in two-phase Ar, in a windowless mode [32].

**Liquid CRAD with cryogenic GPM separated by window from the noble liquid.**

The concept was introduced in 2005 [29] and elaborated in 2011 [30],[31] (see Figs. 11 and 12): the primary scintillation signal produced in noble liquid is detected using a cryogenic GPM with CsI photocathode, separated by window from the liquid.

The proof of principle of this concept was demonstrated:
- first in liquid Xe with cryogenic GPM based on double-THGEM and THGEM/PIM/MM multipliers with CsI photocathode, separated by $MgF_2$ window from the liquid (Fig. 12) [30],[31].

**Liquid Ar CRAD with GAPD optical readout of the THGEM plate.**

The concept was introduced in 2009 [26] and elaborated in 2011 [55],[56] (see Fig. 13): the primary ionization electrons produced in liquid Ar, induce secondary (proportional) scintillations in the holes of a THGEM plate immersed in the liquid, which are optically read out in the VUV using GAPDs with WLS. This concept raises questions, since it employs the idea of noble liquid electroluminescence in hole-type multipliers which in theory requires very high electric fields [57], >1MV/cm, i.e. much higher than those observed in experiment [26].

The proof of principle of this concept was demonstrated (see details in section 4.2):
- first in liquid Ar with GAPD optical readout of the THGEM electroluminescence plate in the VUV using WLS (Fig. 8) [26].

**Two-phase CRAD with Two-Phase Photoelectric Gate.**

The concept was introduced in 2006 [58] (see Fig. 14): the primary ionization electrons produced in noble liquid, induce secondary (proportional) scintillations in the holes of a GEM or MHSP plate immersed in the liquid, which are optically read out in the gas phase using a windowless cryogenic GPM with CsI photocathode. Such a two-phase detector configuration was called "Two-Phase Photoelectric Gate", since it should effectively suppress ion backflow. Similarly to the previous concept, this concept raises questions, since it employs the idea of noble liquid electroluminescence in hole-type multipliers, at electric fields much lower than those expected.

The concept is not fully proved (see details in section 4.2):
- The electroluminescence signal was observed from the GEM plate immersed in liquid Ar, using a cryogenic GPM with CsI photocathode in the gas phase [59].

**Two-phase CRAD with CsI photocathode immersed in the liquid.**

The concept was introduced in 2007 [60] (see Fig. 15): the two-phase Xe CRAD with THGEM multiplier charge readout and CsI photocathode immersed in the liquid was proposed. The concept is based on the results on the CsI photocathode performance in liquid Ar, Kr and Xe [61].

The concept is not proved.

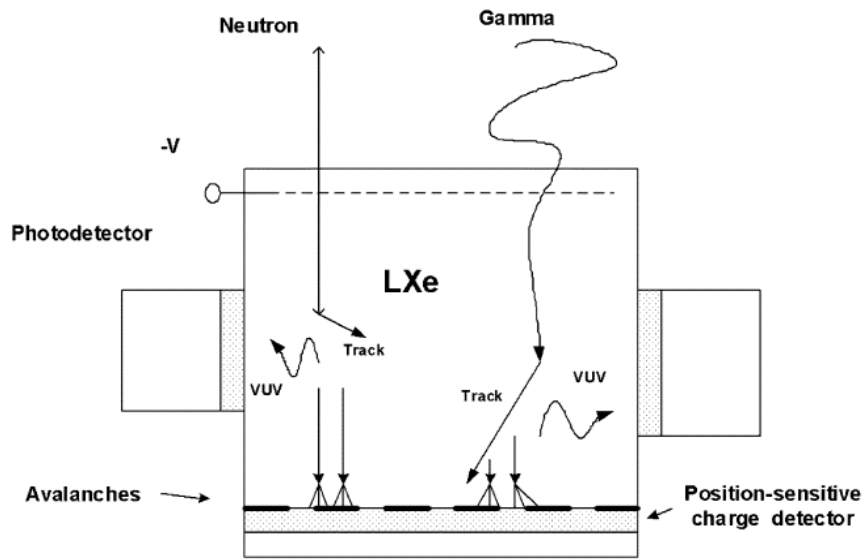

Fig. 11. CRAD with cryogenic Gaseous Photomuliplier (GPM) separated by window from the noble liquid [29].

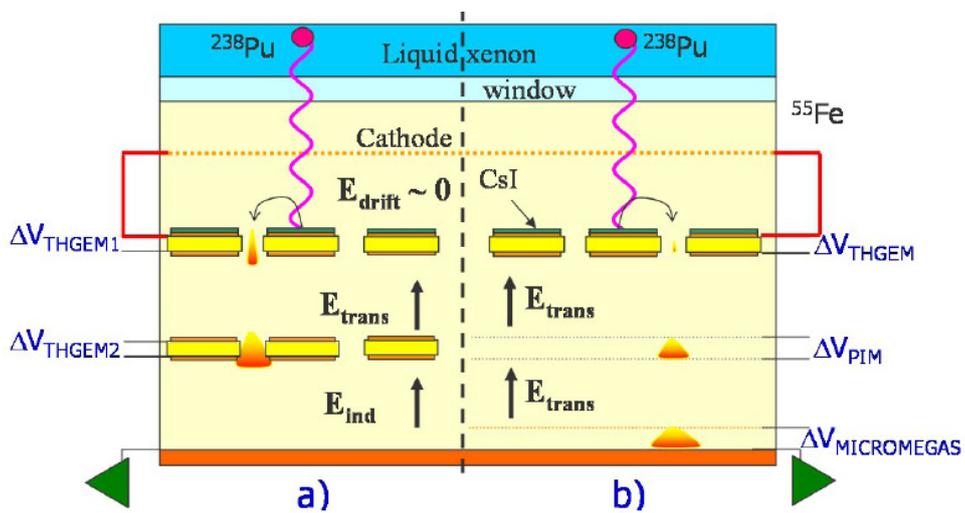

Fig. 12. Liquid Xe CRAD with cryogenic GPM with CsI photocathode based on double-THGEM (a) and THGEM/PIM/MM (b) multipliers, separated by window from the liquid [30],[31].

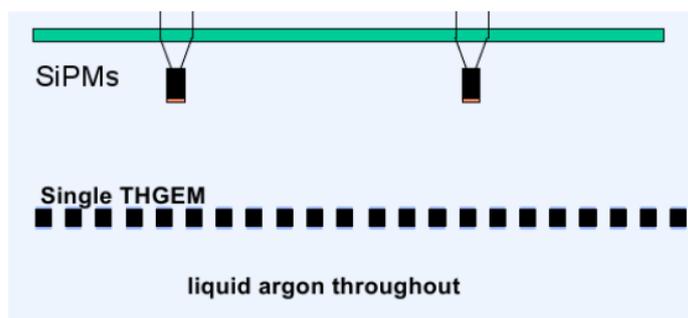

Fig. 13. Liquid Ar CRAD with GAPD (SiPM) optical readout of THGEM plate immersed in the liquid [26],[55],[56].

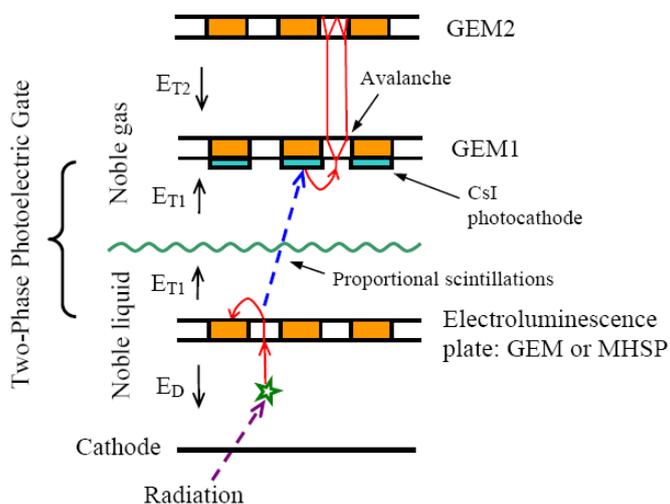

Fig. 14. Two-phase CRAD with Two-Phase Photoelectric Gate, i.e. with optical readout of a GEM or MHSP plate immersed in the liquid, using windowless cryogenic GPM with CsI photocathode [58].

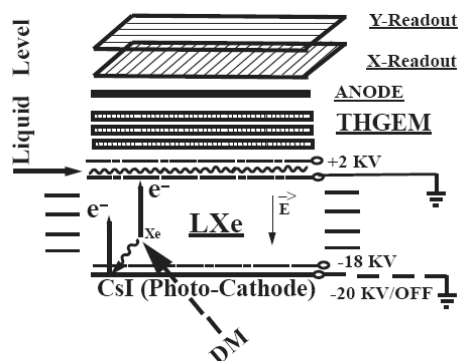

Fig. 15. Two-phase CRAD with CsI photocathode immersed in the liquid and THGEM multiplier charge readout [60].

## 2.4 CRAD-related project concepts

Here we discuss motivations for CRAD developments and possible applications in the field of rare-event experiments and medical imaging techniques. The rare-event applications are characterized by the necessity of detecting signals from nuclei recoils induced by either coherent neutrino-nucleus neutral-current scattering or elastic collision with dark matter particles (WIMPs), on the one hand, and charge-current induced ionization from solar and supernova neutrinos, on the other hand.

Despite the relatively high cross-section, coherent neutrino-nucleus scattering [62] is hard to detect [63],[64],[65] because of the very low energy of recoil nucleus, less than 1 keV, of which only 10-20% goes to ionization . Therefore, the nucleus recoil signal consists of only a few electrons, which necessitates detectors with single-electron sensitivity. For a typical neutrino flux from GW nuclear reactor, the event count-rate is expected to be of several hundreds per day per kg. Accordingly, in coherent neutrino-nucleus scattering the detection threshold and noise rate should be of 1-2 electrons and below $10^{-3}$ Hz per kg of the detection medium respectively; this is a real challenge in detector technology. The use of two-phase CRADs, in Ar and Xe, can provide such sensitivity. The proposed detectors could be considered as "portable" neutrino detectors, with mass of the order of 100 kg. This opens the way for remote control of nuclear reactors by measuring the neutrino flux.

Concerning the dark matter detection, all experiments aim at observing low-energy (0-100 keV) recoils induced by elastic scattering of WIMPs [16],[37]. Similarly to coherent neutrino-nucleus scattering, the signal is weak. With increasing detector scale and mass and lowering background down to ultra-low rates (< 1 event/100 kg/year), replacing the PMTs which currently suffer rather high natural radioactivity, by detectors having lower radioactivity background, could be of high advantage. Therefore, the use of efficient two-phase CRADs in Ar and Xe incorporating advanced MPGDs benefiting of low natural radioactivity, instead of vacuum PMTs, could be an attractive potential solution. Another qualitative step would be the improvement of position accuracy to mm level; in current two-phase detectors with PMTs it is about of a cm which reduces the effective exposure.

Regarding medical imaging, Positron Emission Tomography (PET) is considered to be the most effective means of diagnosing cancer in humans [66],[67]. To ensure early diagnostics, a 511 keV γ-ray detector is needed with higher spatial resolution (~ 1 mm) and better image quality; this is to be compared with existing technique (~ 4 mm), using scintillation crystals and PMTs. Also, the parallax (depth of interaction) problem and that of scattered events are difficult to solve with scintillating crystals. The use of liquid Xe TPCs for PET [16], and in particular liquid Xe CRADs with position-sensitive MPGD readout could address these issues.

The concepts of CRAD-related projects in the field of rare-event experiments and medical imaging techniques are listed below; none of these concepts is fully proven.

**Two-phase Ar detector with THGEM-multiplier charge readout for dark matter search (ArDM project).**

The CRAD-related concept is the following [68],[69] (see Fig. 16): THGEM-based charge readout in the gas phase is combined with PMT-based scintillation readout in the liquid phase, in a two-phase Ar detector, thus employing the concept of the "two-phase CRAD with THGEM-multiplier charge readout" considered in the previous section.

**Giant liquid Ar detector with THGEM-multiplier charge readout for neutrino physics, proton decay and observation of astrophysical neutrinos (GLACIER project).**

The CRAD-related concept is the following [70],[71] (see Fig. 17): the ionization charge attenuated after long drift (>1m) in a giant liquid Ar detector is read out using large-area THGEM multipliers in the gas phase, the detector being operated in the two-phase mode. The concept employs that of the "two-phase CRAD with THGEM-multiplier charge readout" considered in the previous section, operated at moderate charge gains (~100).

**Two-phase Ar and Xe detector with GEM- or THGEM-multiplier charge readout for coherent neutrino-nucleus scattering experiments.**

The CRAD-related concept is the following [65] (see Fig. 18): GEM- or THGEM-based charge readout in the gas phase is combined with PMT-based optical readout of secondary (proportional) scintillations, in a two-phase Ar or Xe detector operated in single electron counting mode. The distinctive feature of the concept is the selection of point-like events having two or more ionization electrons, to reject single-electron background. The concept employs that of the "two-phase CRAD with GEM-multiplier charge readout" considered in the previous section, operated at high gains ($\geq 10^4$).

**Two-phase Ar detector with THGEM/GAPD-matrix optical readout in the NIR for coherent neutrino-nucleus scattering and dark matter search experiments.**

The CRAD-related concept is the following [35],[36] (see Fig. 19): THGEM/GAPD-matrix optical readout (in the NIR) of the charge signal in the gas phase, is combined with PMT-based optical readout of secondary (proportional) scintillations in the gas phase and of primary scintillations in the liquid phase, if any, in a two-phase Ar detector operated in single-electron counting mode. The secondary (proportional) scintillations in the gas phase, recorded in the VUV using PMTs, provide a single-photoelectron trigger due to the excellent amplitude resolution available in the proportional scintillation mode. The avalanche scintillations, recorded in the NIR using a matrix of GAPDs, provide a good spatial resolution. Such an optical readout is preferable as compared to charge readout in terms of overall gain and noise. The concept employs that of the "CRAD with optical readout using combined THGEM/GAPD multiplier" considered in the previous section, operated at high gains.

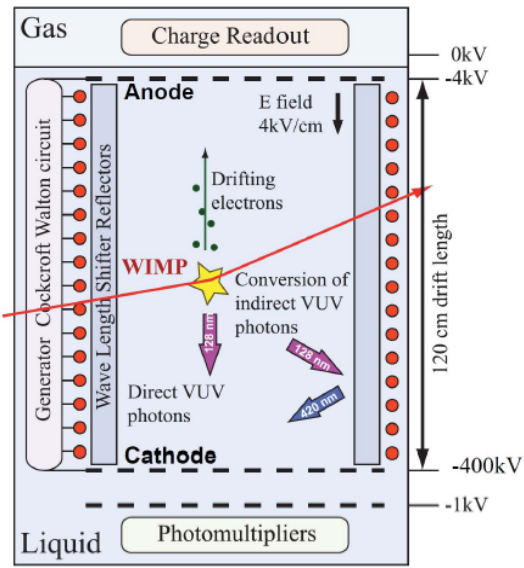

Fig. 16. Two-phase Ar detector with THGEM-multiplier charge readout for dark matter search (ArDM project) [68],[69].

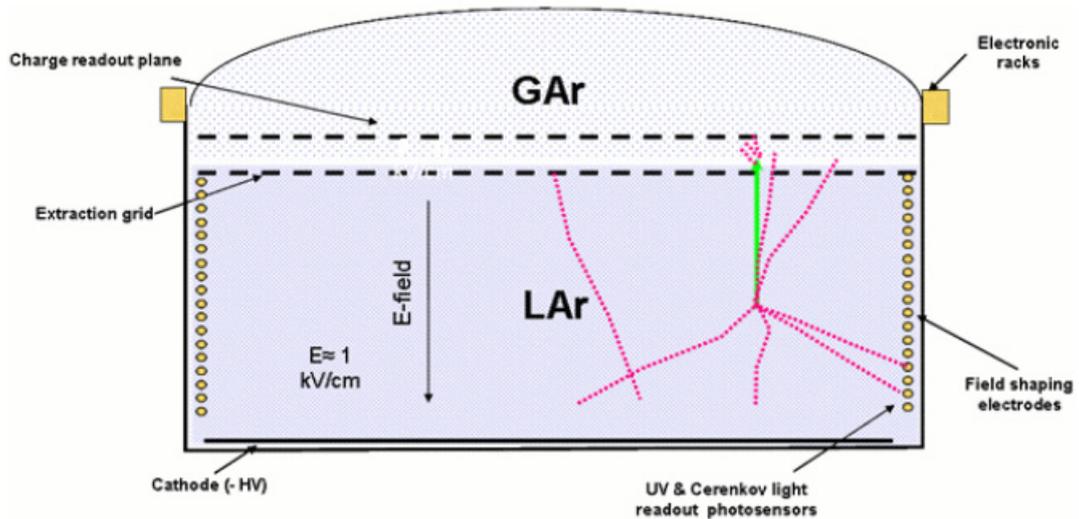

Fig. 17. Giant liquid Ar detector for neutrino physics, proton decay and observation of astrophysical neutrinos (GLACIER project) [70],[71].

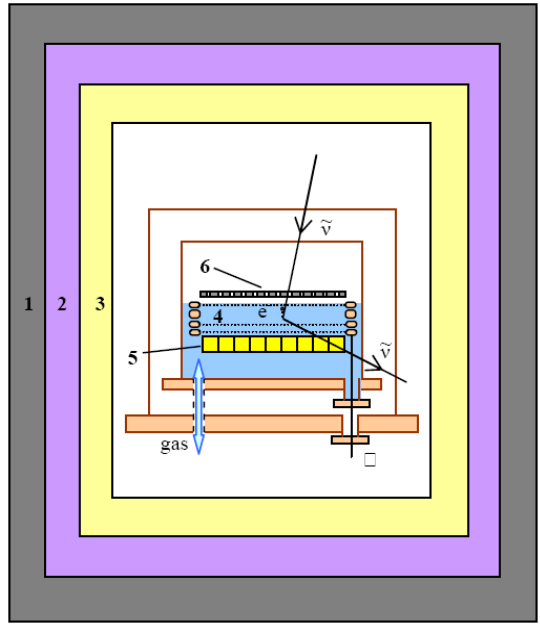

Fig. 18. Two-phase Ar and Xe detector with GEM- or THGEM-multiplier charge readout for coherent neutrino-nucleus scattering experiments [65]. 1 - lead shield, 2 – plastic scintillator, 3 - Gd-loaded plastic, 4 – noble liquid, 5 - PMTs, 6 – GEM or THGEM.

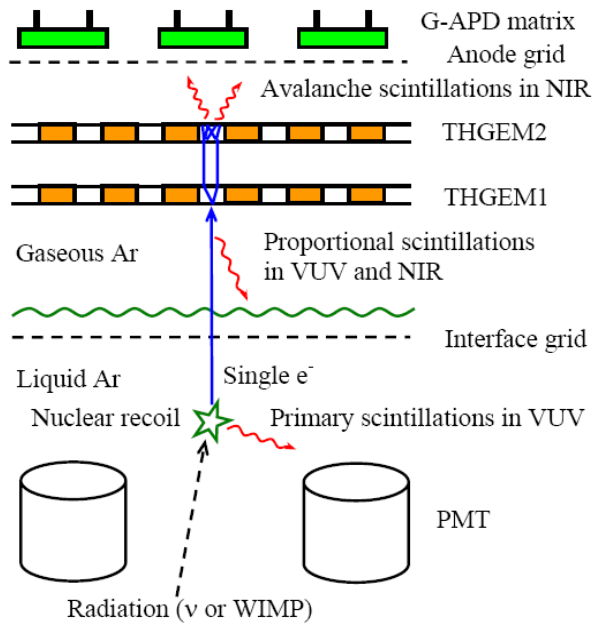

Fig. 19. Two-phase Ar detector with THGEM/GAPD-matrix optical readout in the NIR for coherent neutrino-nucleus scattering and dark matter search experiments [36].

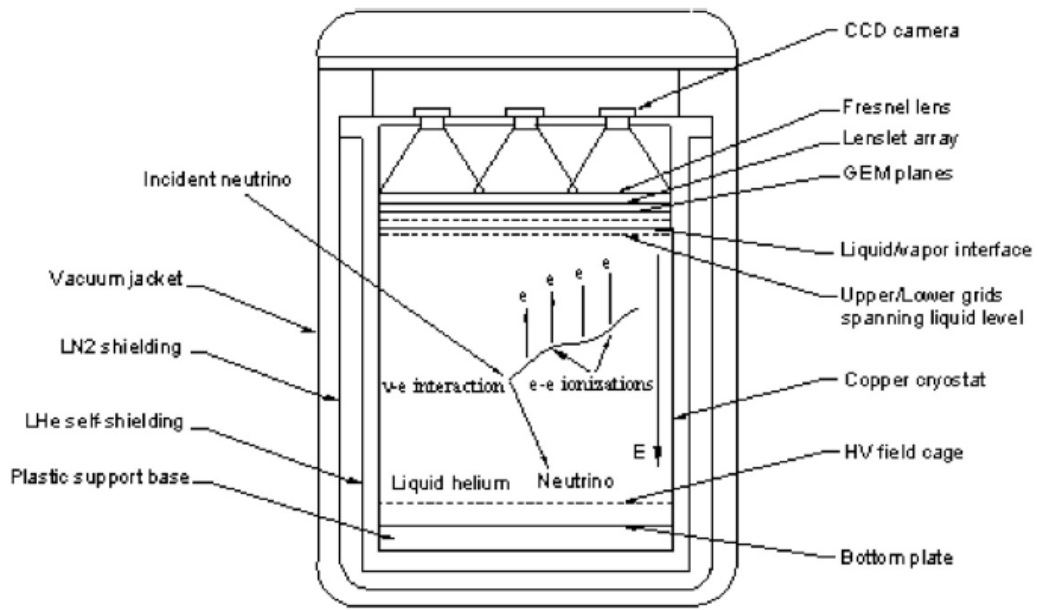

Fig. 20. Two-phase He detector with GEM/CCD optical readout for solar neutrino detection (E-bubble project) [33].

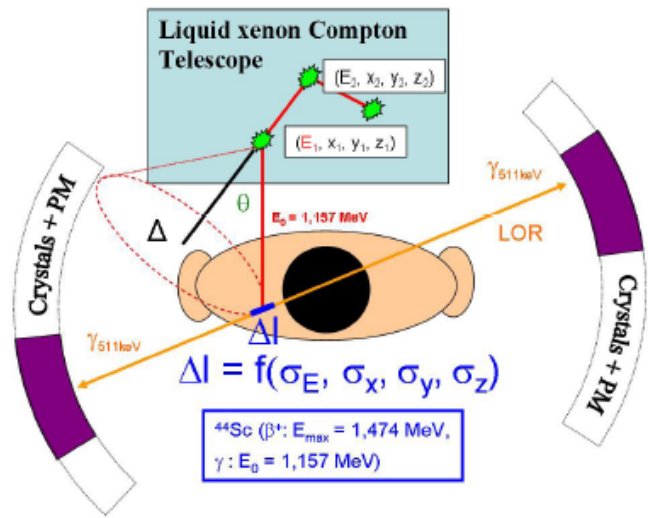

Fig. 21. 3γ-PET with Compton telescope based on liquid Xe CRAD with cryogenic GPM [73],[74].

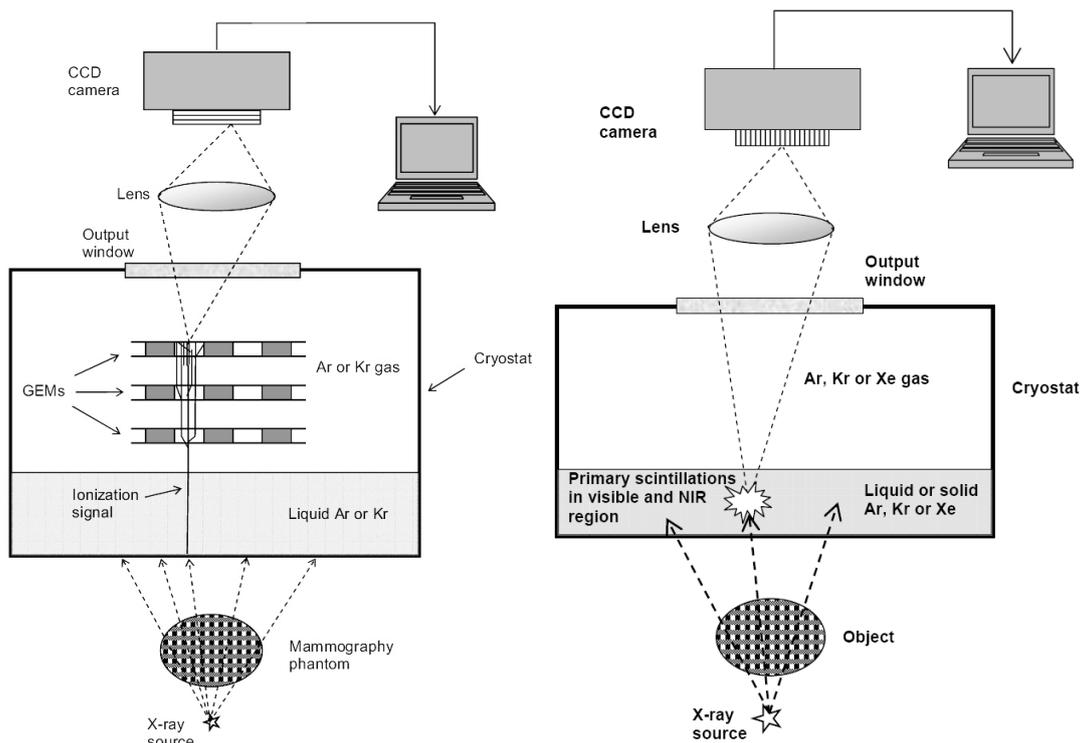

Fig. 22. Two-phase detectors with GEM/CCD optical readout in the NIR for digital mammography (left) and with CCD optical readout in the visible and NIR region for digital radiography (right).

**Two-phase or high-pressure He and Ne detectors with GEM/CCD optical readout for solar neutrino detection (E-bubble project).**

The CRAD-related concept is the following [33],[34] (see Fig. 20): the ionization produced by neutrino charge-current scattering in either a two-phase He or high-pressure Ne cryogenic detector, is recorded in the gas phase using CCD-camera optical readout from the GEM multiplier. The distinctive feature of the concept is that the ionization in liquid He or supercritical high pressure Ne at low temperatures is localized in electron bubbles, having low diffusion and thus providing high spatial resolution. This results in superior track imaging capability and possibility to reconstruct the direction of the incident solar neutrino, which considerably enhances the effective exposure.

**Two-phase Xe detector with GEM- or THGEM-multiplier charge readout for PET.**

The CRAD-related concept is the following [12],[72]: a two-phase Xe CRAD with GEM- or THGEM-based charge readout is used for detection of back-to-back 511 keV γ-rays, in 3D readout mode. This will help to solve the parallax (depth of interaction) problem and obtain the superior (~1 mm) spatial resolution. In addition, due to the high granularity of GEM-based readout one can measure Compton double scattering and thus determine the angle of the γ-ray: this will help to solve the problem of scatter and random events, and thus dramatically improve the image contrast and quality.

### 3γ-PET with Compton telescope based on liquid Xe CRAD with cryogenic GPM.

The CRAD-related concept is the following [73],[74] (see Fig. 21). In a standard PET device the $\beta^+$ emitter localization is based on the detection in coincidence of the two back-to-back γ-rays. Hence, the position of the emitter is known only along the line-of-response. Using a specific radioisotope emitting a γ just after the $\beta^+$ decay permits a detection of the three photons. The direction of the additional emitted γ-ray is measured with a Compton telescope, based on the liquid Xe CRAD with cryogenic GPM [30],[31] considered in the previous section. The position of the emitter can then be measured by calculating the intersection with the line-of-response.

### Two-phase detector with GEM/CCD or CCD optical readout for digital radiography.

The CRAD-related concept is the following (see Fig. 22). The first idea is to use a two-phase (liquid-gas) Ar or Kr CRAD with GEM/CCD optical readout in the NIR, for digital mammography (Fig. 22, left). Due to the small absorption length and photoelectron range for soft (≤40 keV) X-rays, below 500 and 20 microns in liquid Kr respectively, the liquid or solid layer thickness can be done as small as a few mm. This will help to avoid the parallax problem.

The second idea is to use a two-phase CRAD in liquid-gas or solid-gas state with direct CCD readout of primary scintillations in the noble liquid (or solid), in the visible and NIR regions, for digital radiography (Fig. 22, right). This idea is based on the fact that primary scintillations in liquid Ar has noticeable yield [36], of the order of 500 photon/Mev in the visible and NIR range where CCDs have high sensitivity.

In conclusion to this section, the following types of detectors need to be developed:
- high-gain and low-noise two-phase Ar and Xe CRADs having single-electron sensitivity, for coherent neutrino-nucleus scattering and dark matter search experiments;
- large-area moderate-gain two-phase CRADs in Ar, for giant liquid Ar detectors, and in Xe, for PET;
- liquid Xe CRADs with high-gain cryogenic GPMs with CsI photocathode, for PET.

The R&D results in these directions will be presented in the next chapter.

## 3. CRAD R&D results

Over the past 8 years there has been an intense and difficult R&D work of different groups in the field of CRAD developments. As an example of this work, Fig. 23 shows the latest version of the experimental setup of the Budker INP group to study gaseous and two-phase CRADs in Ar and Xe with charge and optical readout [28],[35]: it comprises a 9 l cryogenic chamber containing an assembly of the combined THGEM/GAPD multiplier.

In this chapter the most important results of this R&D work are presented.

## 3.1 Gaseous CRADs

Gaseous CRADs (Fig. 1, left) are in fact high-pressure detectors, the operation of which at cryogenic temperatures allows to significantly increase the density of the medium, because the gas density is approximately inversely proportional to the temperature.

It should be remarked that the stable operation of GEMs and other hole-type MPGDs at low temperatures is a non-trivial fact, since the electrical resistance of their dielectric materials considerably increases with the temperature decrease, i.e. by an order of magnitude per 35 degrees for Kapton GEMs [19], which might result in strong charging-up effects within the holes. Fortunately the latter did not happen: the charging-up effects have not been observed; one can see this from Fig. 24 demonstrating the independence of the triple-GEM gain characteristic of the primary ionization flux at cryogenic temperature.

The GEM performance in gaseous CRADs was found to be generally independent of temperature at cryogenic temperatures down to ~100 K [1],[19]: stable and high-gain GEM operation was observed in all noble gases and in their mixtures with selected molecular additives that do not freeze in a wide temperature range ($CH_4$, $N_2$ and $H_2$). This is seen from Fig. 25 [12]: rather high triple-GEM gains were reached at cryogenic temperatures, exceeding $10^5$ in He and $10^4$ in Ar, Kr and Xe+$CH_4$.

In these and in the following measurements the typical GEM geometrical parameters were the following: dielectric (Kapton) thickness was 50 μm, hole pitch - 140 μm, hole diameter on metal - 70 μm.

As concerns the THGEM performance in gaseous and two-phase CRADs, the THGEM is thicker and has over an order of magnitude fewer holes than a GEM; accordingly, it is expected to be more robust, with better resistance to discharges. Similarly to GEMs, the THGEM and RETHGEM multiplier performances in electron avalanching mode were stable at cryogenic temperatures [28],[44],[45],[75], with gains exceeding $10^3$ in Ar [75] and reaching 600 in Xe [28] at gas densities corresponding to those of saturated vapour in the two-phase mode: see Fig. 26.

It is amazing that GEMs were able to operate in electron avalanching mode at even lower temperatures, down to 2.6 K in gaseous He [20]. On the other hand, high GEM gains observed in He and Ne above 77 K were reported to be due to the Penning effect in uncontrolled ($\geq 10^{-5}$) impurities (i.e. $N_2$) which froze out at lower temperatures, resulting in the considerable gain drop at temperatures below 40 K [20],[21]: see Fig. 27 (left). In more detail electron avalanching mechanisms at low temperatures will be discussed in section 4.4. A solution to the gain drop problem at lower temperatures was found in ref. [20]: Ne and He can form high-gain Penning mixtures with $H_2$ at temperatures down to ~10 K. This is seen from Fig. 27 (right): triple-GEM gains exceeding $10^4$ were obtained at 57 K in the Penning mixture Ne+0.1%$H_2$, its density corresponding to that of saturated Ne vapour in the two-phase mode. Unfortunately, this does not work for two-phase He, due to the very low $H_2$ vapour pressure at 4.2 K. Nevertheless, this solution permitted to observe α-particle tracks in dense Ne at 77 K, using CCD optical readout of single-GEM multiplier [34] (see Fig. 28), thus proving to some extent the concept of the "high-pressure Ne detector with GEM/CCD optical readout for solar neutrino detection" [33],[34] considered in section 2.4.

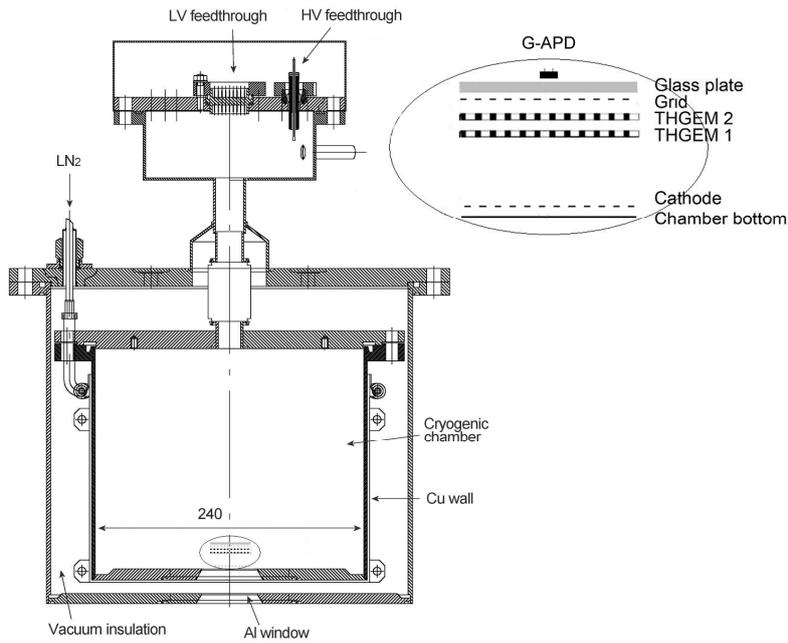

Fig. 23. Design drawing (to scale) of the cryogenic chamber to study gaseous and two-phase CRAD performances in Ar and Xe, with combined THGEM/GAPD multiplier readout; in the insert is an expanded view of the double-THGEM/GAPD assembly mounted inside [28],[35].

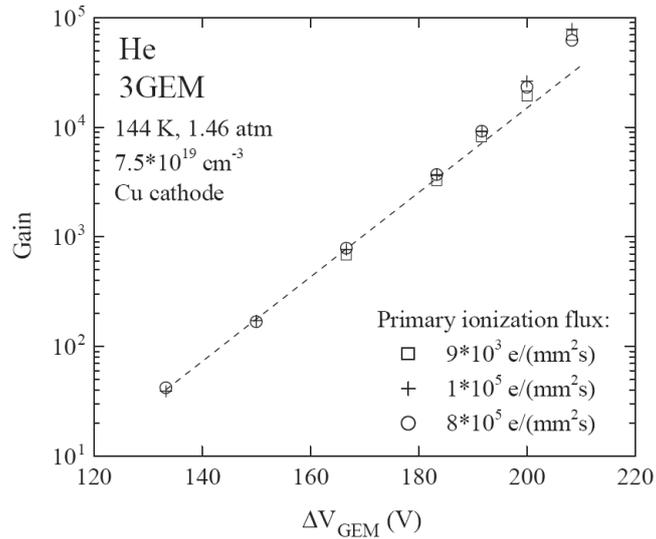

Fig. 24. Gain characteristics of a triple-GEM multiplier in gaseous He at 144 K and 1.46 atm, at different fluxes of the primary ionization induced by X-rays [19].

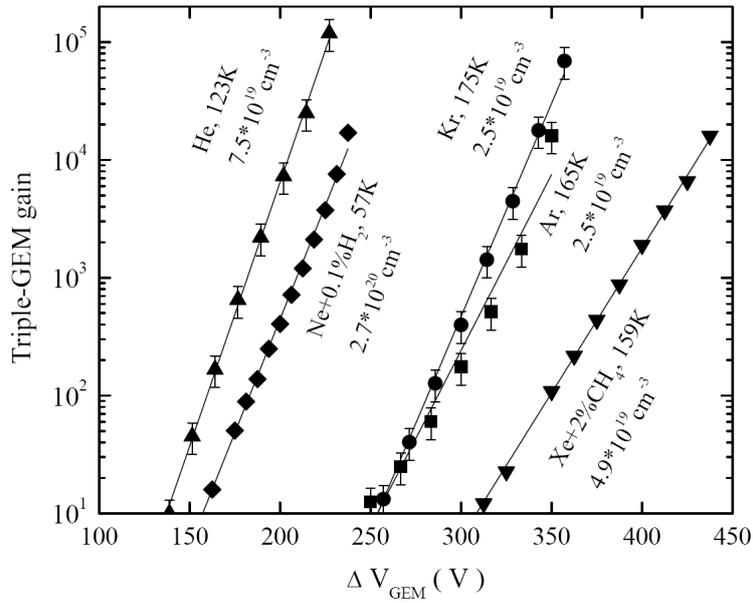

Fig. 25. Gain characteristics of triple-GEM multipliers at cryogenic temperatures in noble gases and their mixtures with molecular additives [12]. The appropriate temperatures and atomic densities are indicated.

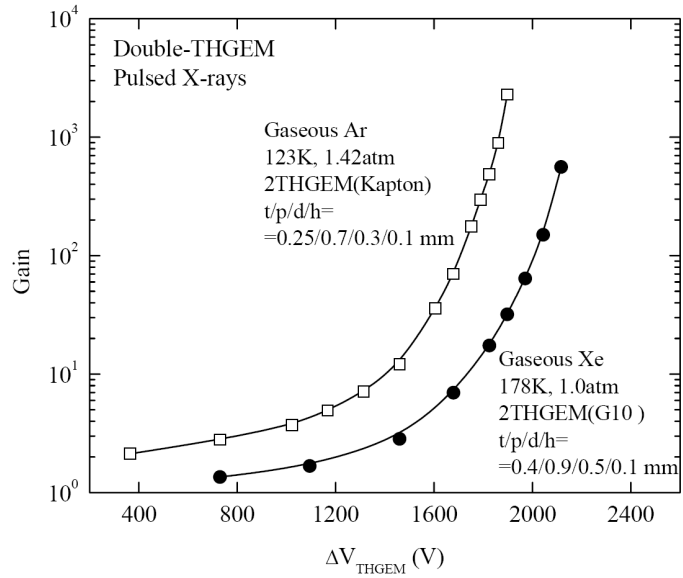

Fig. 26. Gain characteristics of double-THGEM multipliers at cryogenic temperatures, in gaseous Ar with THGEM made of Kapton [75] and in gaseous Xe with THGEM made of G10 [28], at gas densities corresponding to those of saturated vapour in the two-phase mode. The maximum gains are limited by discharges. The appropriate temperatures, pressures and THGEM geometrical parameters are indicated. Here t/p/d/h denotes "dielectric thickness/hole pitch/hole diameter/hole rim width".

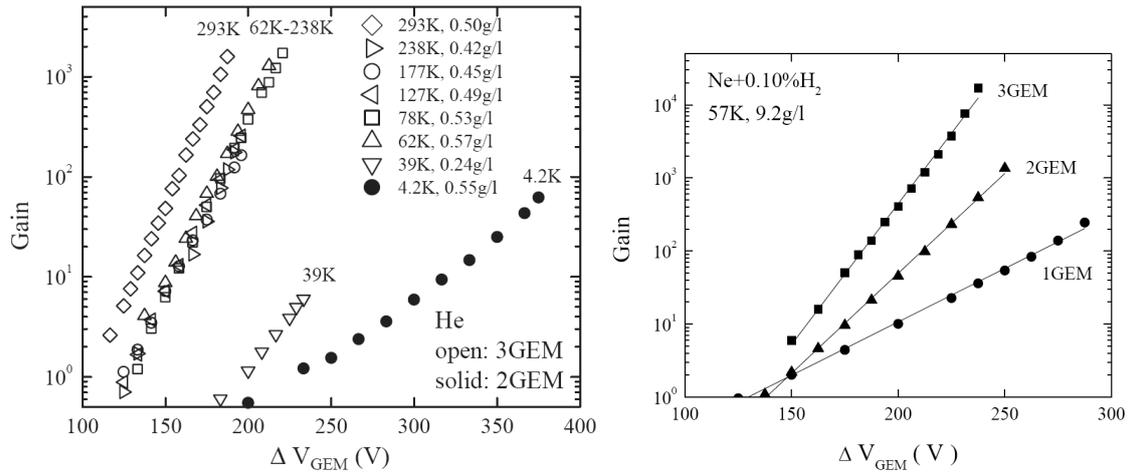

Fig. 27. Gain characteristics of GEM multipliers at low temperatures (down to 4.2 K) in gaseous He (left) and in the Penning mixture Ne+0.1%$H_2$, its density corresponding to saturated Ne vapour in the two-phase mode (right) [20]. The appropriate temperatures and gas densities are indicated. At 39 K and 4.2 K (left), the maximum gains were limited by discharges, while at other temperatures the discharge limit was not reached.

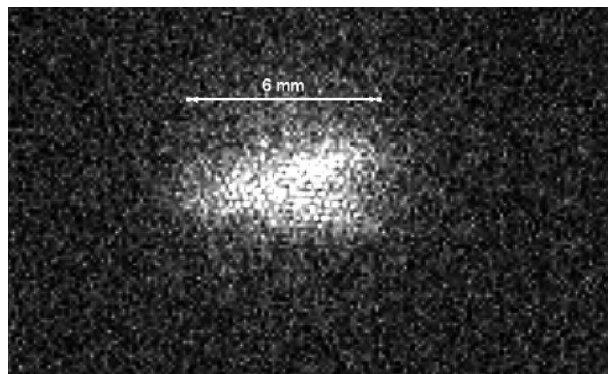

Fig. 28. Two α-particle tracks image obtained in gaseous CRAD with GEM/CCD optical readout in dense (22 g/l) Penning mixture Ne+0.1%$H_2$ at 77 K, at single-GEM gain in excess of 1000 [34]. Here the track width is dominated by the coulomb spread of charge.

## 3.2 Two-phase CRADs with charge readout

The operation principles of two-phase emission detectors, i.e. of those with electron emission through the liquid-gas interface, were studied several tens years ago [13],[16],[17]. The operation of such detectors in electron avalanching mode, using wire chamber readout, turned out to be unstable [8]. Accordingly, the conventional two-phase detectors used so far in dark matter search experiments operate without electron avalanching, with optical readout using PMTs [16],[37]. It was the discovery of the high-gain operation of GEM multipliers in noble gases [11],[18],[39], and in particular in the two-phase mode [1],[22], that changed the situation and substantially advanced two-phase detectors with charge readout (i.e. two-phase CRADs).

In two-phase CRADs with charge readout considered in sections 2.3 (Figs. 1 and 7) and 2.4 (Figs. 16-18), the preferable detection media are Ar and Xe. They have the highest cross-sections for nuclear recoils induced by weakly interacting particles, such as neutrino or WIMPs. Kr is excluded here due to its natural radioactivity, though it might be used in digital radiography projects (Fig. 22), while Xe might be very attractive for medical applications, such as PET (Fig. 21), due to its higher Z (see section 2.4).

Most promising results have been obtained with two-phase Ar CRADs (see Fig. 1): charge gains of $10^4$ with triple-GEM readout were routinely attained [22],[32],[53]; the stable operation for tens of hours at gains two-fold lower than the maximum was demonstrated [53]. These maximum gain values should be compared to those of 600 and 200 obtained in GEM-based two-phase Kr and Xe CRADs respectively: see Fig. 29 [22].

In this review the CRAD charge gain is defined as that of the Budker INP and Weizmann Institute groups [28]: it is the ratio of the output anode charge of the MPGD multiplier incorporated into the CRAD to the input "primary" charge, i.e. to that of prior to multiplication measured in special calibration runs. The anode signals are read out from either the last electrode of the last GEM, i.e. in a 1GEM, 2GEM or 3GEM operation mode (as shown for example in Fig. 9), or the Printed Circuit Board (PCB) electrode behind the last GEM, i.e. in a 1GEM+PCB, 2GEM+PCB or 3GEM+PCB mode. Similar notations are used here for THGEM multiplier operation modes. It should be remarked that in the GEM+PCB mode the charge gain typically amounts to 1/3-1/2 of that in the GEM mode, since the avalanche charge is not fully transferred to the PCB from the GEM holes [12],[76]: some its fraction is collected at the GEM electrode. On the other hand, the GEM+PCB mode is more practical when the 1D or 2D position sensitivity is required: the PCB can be appropriately patterned with readout strips.

Relatively high GEM gains provided a wide dynamical range of two-phase Ar CRADs, permitting to effectively discriminate signals induced by single electrons, elastically scattered neutrons and 60 keV X-rays: this is seen from Fig. 30 showing the appropriate amplitude spectra [53]. From this figure it is seen that the energy resolution of two-phase Ar CRADs is not that good: for 60 keV X-rays it amounted to $\sigma/A=17\%$ which is worse than that of two-phase emission detectors based on PMTs. This is also seen from Fig. 31 demonstrating single-electron amplitude spectra obtained in two-phase Ar CRADs using external trigger [52]: though the spectra are well separated from electronic noise at gains exceeding ~5000, they are described by an exponential function (rather than by a peaked function). The latter is generally a rule for gaseous multipliers operated in proportional mode, due to intrinsically considerable fluctuations of the avalanche size [77]. Consequently, the single- and double-electron events can hardly be distinguished in two-phase Ar CRADs with GEM (or THGEM) charge readout. For that the combination with PMT-based optical readout should be used, as suggested in section 2.4 when discussing the CRAD concepts related to coherent neutrino-nucleus scattering projects. Here the function of the GEM- or THGEM-based charge readout is to provide superior spatial resolution and low noise.

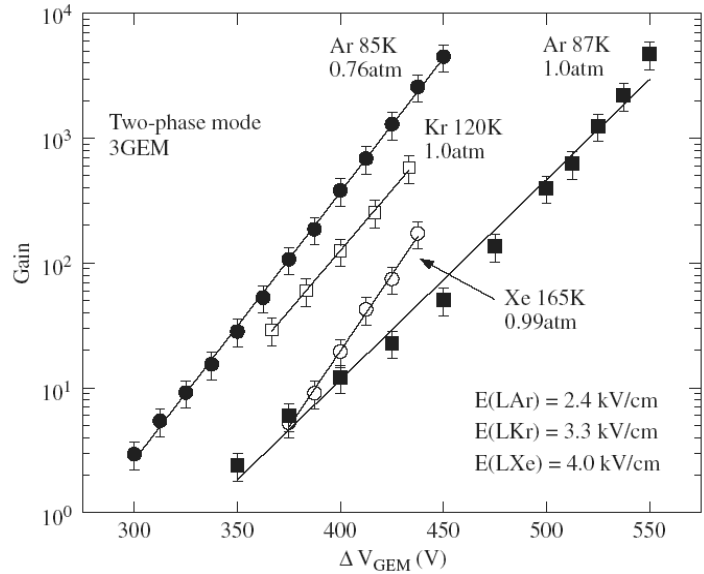

Fig. 29. Gain characteristics in two-phase Ar and Xe CRADs with triple-GEM multiplier charge readout [22]. The appropriate temperatures, pressures and electric fields in the liquids are indicated. The maximum gains are limited by discharges.

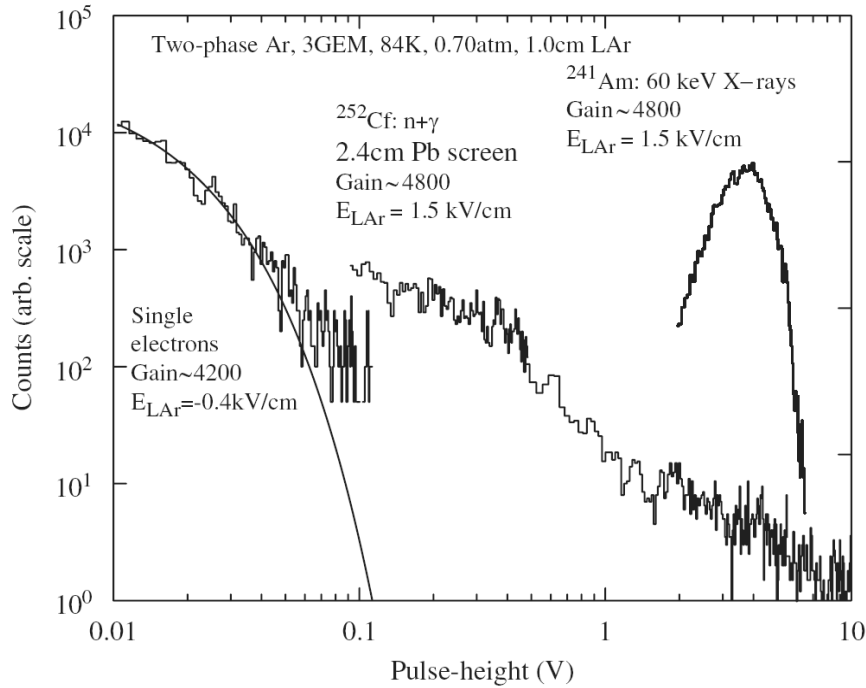

Fig. 30. Pulse-height spectra in two-phase Ar CRADs with triple-GEM multiplier charge readout, induced by single electrons, neutrons and $\gamma$-rays from $^{252}$Cf source and 60 keV X-rays from $^{241}$Am source, at gains in the range of 4000–5000 [53]. The single-electron spectrum is fitted by an exponential function.

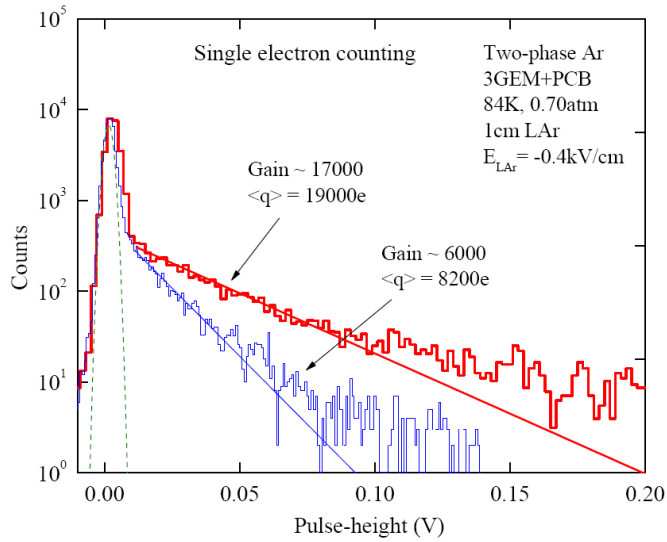

Fig. 31. Pulse-height spectra in a two-phase Ar CRAD with triple-GEM multiplier charge readout (in 3GEM+PCB mode), operated in a single electron counting mode, at gains of 6000 and 17000, obtained with external trigger [52]. An electronic noise spectrum (dashed line) is also shown.

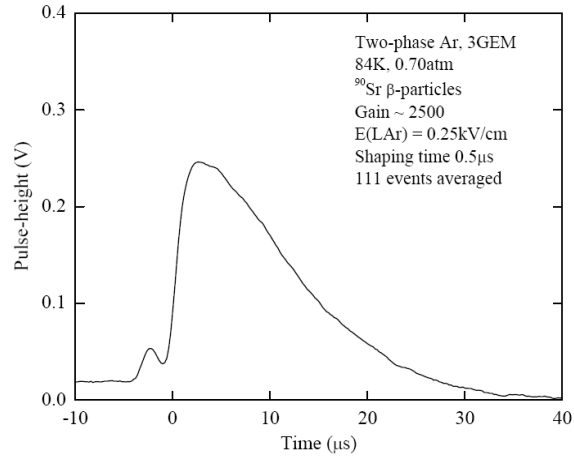

Fig. 32. Average anode signal in a two-phase Ar CRAD with triple-GEM multiplier readout of both the ionization and primary scintillation signals, using CsI photocathode on the first GEM, at a charge gain of 2500 and electric field in the liquid of 0.25 kV/cm [32]. The average energy deposited by β-particles in the active liquid layer was 600 keV. The primary scintillation (the first) and ionization (the second) signals are distinctly seen.

As was discussed in section 2.3, the GEM multiplier can provide the detection of both the charge (ionization) signal and that of primary scintillations, by depositing a CsI photocathode on the first GEM (Fig. 6). The scintillation signal could provide the trigger to readout the ionization signal and to measure the position in depth, like in TPC. It might be also used to select the useful nuclear-recoil events in coherent neutrino-nucleus scattering and dark matter search experiments, by comparing to the ionization signal. In addition, the fast scintillation signal in liquid Xe could provide the trigger for coincidences of two back-to-back γ-rays in PET. The realization of this concept however had a limited success [32],[53],[72]: the primary scintillation signal was indeed observed along with that of ionization in two-phase Ar CRADs with triple-GEM multiplier readout and CsI photocathode on the first GEM (see Fig. 32). However, the amplitude of the scintillation signal was rather small, of only 30 photoelectrons per 600 keV energy deposited in the liquid [32], presumably due to combined effects of photoelectron backscattering in dense noble gas and poor photoelectron collection into the GEM holes. This amplitude value is not sufficient for rare-event experiments [53], but is enough for PET if operated at charge gains in excess of 5000 [72]. The latter however still has to be demonstrated for two-phase Xe CRADs, which is a difficult task in particular in view of lower charge gains obtained in Xe compared to Ar.

Regarding two-phase CRADs with THGEM multiplier charge readout (see Fig. 7), their maximum gains are comparable to those with GEM multiplier readout: see Figs. 33 and 34. Gains as high as 3000 [24] and 600 [28] were obtained in two-phase CRADs in Ar and Xe respectively, with double-THGEM multipliers having active area of 2.5×2.5 cm$^2$. Their typical anode signals in Ar are shown in Fig. 35, induced by 1000 and 50 primary electrons (prior to multiplication) [24]. The relatively large pulse width, of a few tens of μs, is explained by the joint action of two effects: that of the slow component of electron emission through the Ar liquid-gas interface [78] and that of the experimental fact that the anode signal in the THGEM operated at dense Ar is inherently slower at higher gains [24],[35].

In these measurements the typical THGEM geometrical parameters were the following: t/p/d/h=0.4/0.9/0.5/0.1 mm. Here t/p/d/h denotes "dielectric thickness/hole pitch/hole diameter/hole rim width". The copper layer thickness on THGEM electrodes was typically 30 μm.

A relatively high gain, of about 80, was obtained in a two-phase Ar CRAD with the single-THGEM of larger active area, of 10×10 cm$^2$, having 2D readout [27]: see Fig. 36. The 2D readout permitted to obtain track images (Fig. 36), demonstrating the excellent imaging capability of two-phase Ar CRADs with THGEM multiplier charge readout even at such a moderate gain, which is of particular importance for giant liquid Ar detectors considered in section 2.4. Higher gains, reaching 1000, have been recently obtained in a two-phase Ar CRAD with the double-THGEM of 10×10 cm$^2$ active area: see Fig. 37 and ref. [75].

Noise-rates were assessed in two-phase Ar CRADs with both GEM and THGEM multipliers [24]. At a detection threshold of 4 primary electrons the noise rate of the GEM multiplier was about 0.2 Hz per 1 cm$^2$ of the detector's active area. At a threshold of 20 primary electrons and with pulse-shape analysis, the noise rate of the THGEM multiplier was as low as 0.007 Hz per 1 cm$^2$. These results pave the way towards a new generation of "noiseless" detectors, with noise rates below $10^{-3}$ Hz per kg, as requested in coherent neutrino-nucleus scattering and other rare-event experiments.

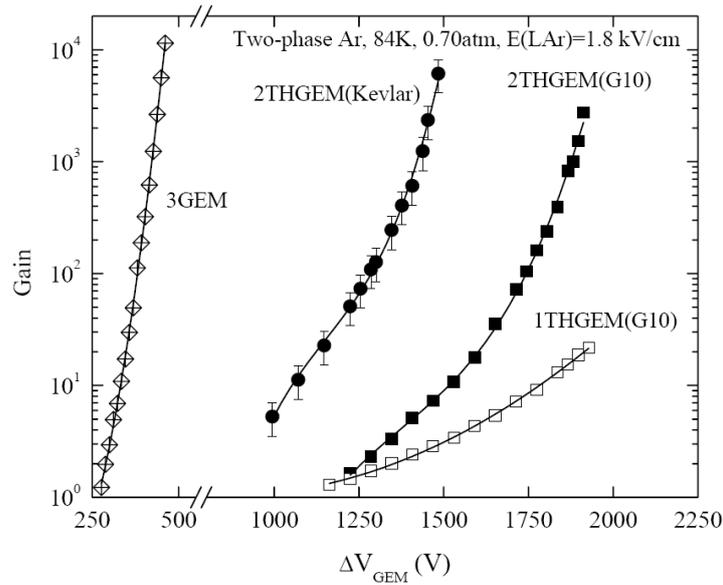

Fig. 33. Gain characteristics in two-phase Ar CRADs with single- and double-THGEM(G10) and double-THGEM(Kevlar) multipliers charge readout [24]. For comparison, that with triple-GEM multiplier is shown. The maximum gains were limited by discharges (except of that in the single-THGEM). The THGEM active area was 2.5×2.5 cm$^2$. The THGEM(G10) geometrical parameters are t/p/d/h=0.4/0.9/0.5/0.1 mm.

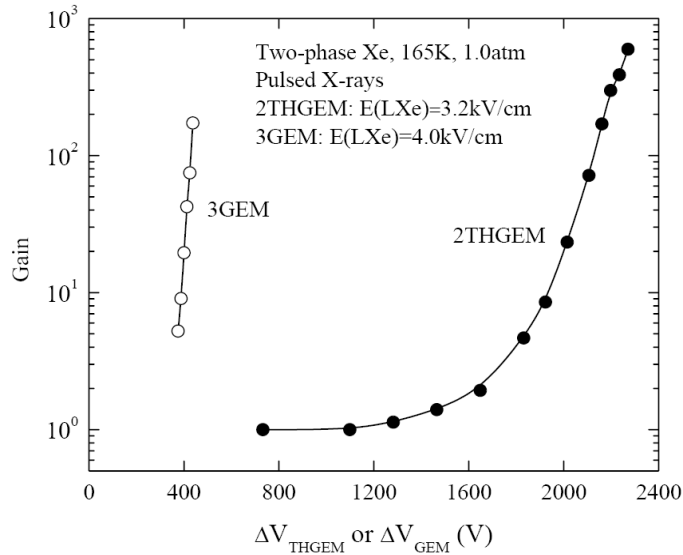

Fig. 34. Gain characteristic in a two-phase Xe CRAD with double-THGEM(G10) multiplier charge readout [28]. For comparison, that with triple-GEM multiplier readout is shown. The maximum gains were limited by discharges. The THGEM active area was 2.5×2.5 cm$^2$. The THGEM geometrical parameters are t/p/d/h=0.4/0.9/0.5/0.1 mm.

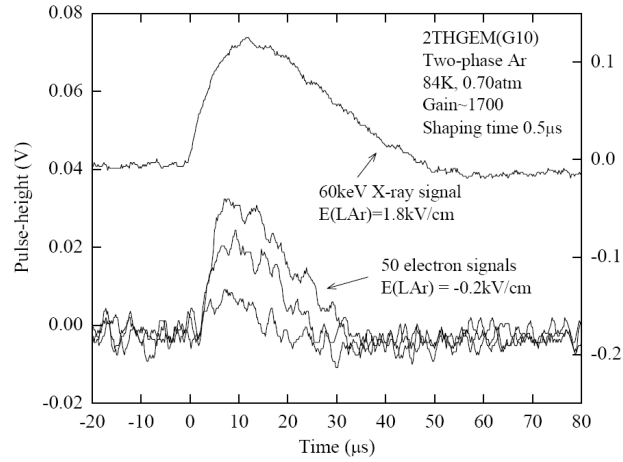

Fig. 35. Typical anode signals in a two-phase Ar CRAD with double-THGEM multiplier charge readout in 2THGEM mode [24]. Pulses were induced by a 60 keV X-ray from $^{241}$Am producing ~1000 primary (prior to multiplication) electrons (top, right scale), and by pulsed X-rays producing ~50 primary electrons (bottom, left scale). Detector charge gain was 1700.

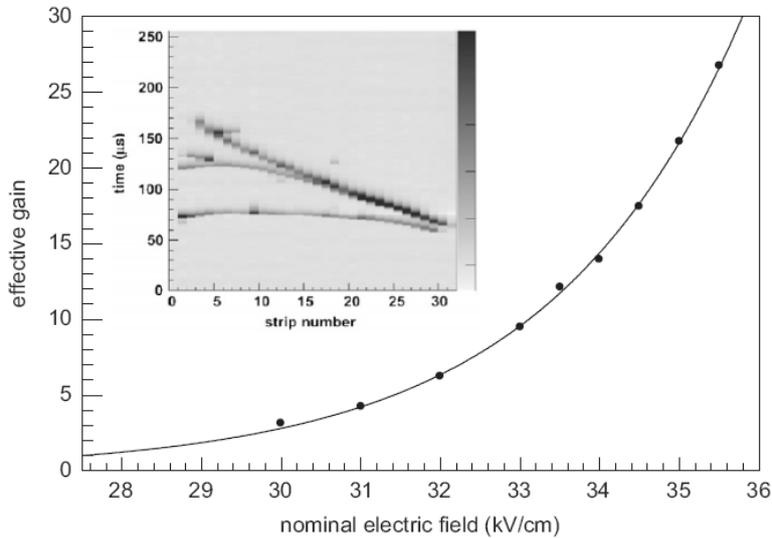

Fig. 36. "Effective" gain characteristic in a two-phase Ar CRAD with single-THGEM multiplier charge 2D readout in 1THGEM+PCB operation mode, having 10×10 cm$^2$ active area [27]. In the insert is an image of the reconstructed muon tracks. The "effective" gain value should be multiplied by a factor of 3, to be normalized to the gain definition of the present review [28]. The THGEM geometrical parameters are t/p/d/h=1/0.8/0.5/0.05 mm.

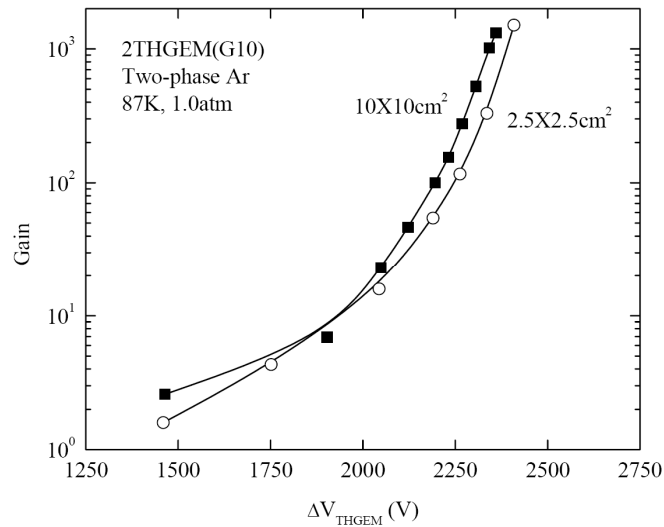

Fig. 37. Gain characteristic in a two-phase Ar CRAD with double-THGEM multiplier charge readout in 2THGEM operation mode, having 10×10 cm$^2$ active area [75]. For comparison that of 2.5×2.5 cm$^2$ active area is shown. The THGEM geometrical parameters are t/p/d/h=0.4/0.9/0.5/0.1 mm.

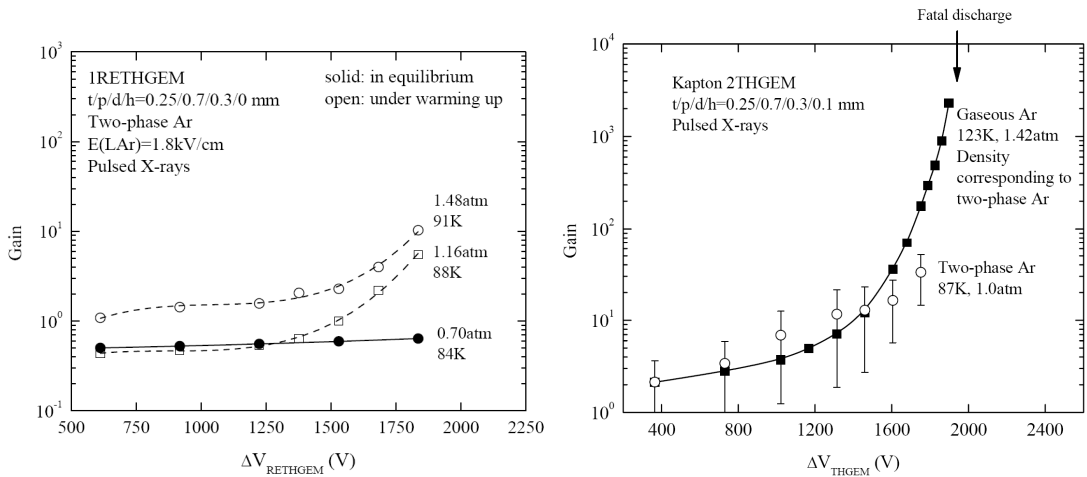

Fig. 38. Gain instabilities observed in two-phase Ar CRADs with RETHGEM (left, [24]) and Kapton THGEM (right, [75]) multipliers charge readout. Shown are the gain characteristics of the single-RETHGEM in two-phase Ar in equilibrium and under warming-up conditions (left), and that of the Kapton double-THGEM in two-phase Ar (right); for the latter that in gaseous Ar is also shown at gas density corresponding to that of saturated vapour in the two-phase mode. Here gain instabilities mean either non-multiplication in the equilibrium state (left) or large gain variations represented by the error bars (right). The RETHGEM and THGEM geometrical parameters are indicated in the figures.

Recently, first results on the two-phase Ar CRAD performance with MPGD multiplier readout other than that of hole-type, namely with that of Micromegas (MM) multiplier, have been reported though with rather low maximum gains, of the order of 5 [79].

We found it useful to summarize in Table 1 all presently existing data on maximum gains attained in two-phase Ar, Kr and Xe CRADs incorporating MPGD multipliers. The following conclusions can be drawn from Table 1:

1. In a sequence "Ar, Kr, Xe" the maximum gain of two-phase CRADs decreased from Ar to Xe by more than an order of magnitude and by half an order of magnitude for GEM and THGEM multipliers respectively.

2. In terms of the maximum reachable gain in two-phase CRADs, the most efficient were Ar-operated ones: the maximum gain reached values of several thousands, both with GEM and THGEM multipliers. These should be compared to values of the order of 500 in two-phase Xe CRADs.

3. The data obtained by different groups on the maximum gains are in fair agreement.

| Group | Two-phase medium | Multiplier type and operation mode | Active area, cm | Typical maximum gain | Reference |
|---|---|---|---|---|---|
| Budker INP | Ar | 3GEM | 2.8×2.8 | $(5-10)\times 10^3$ | Bondar et al, NIM A 556 (2006) 273, 598 (2009) 121 |
| Budker INP, Weizmann Inst. | Ar | 2THGEM | 2.5×2.5 | 3000 | Bondar et al, JINST 3 (2008) P07001 |
| Budker INP, Weizmann Inst. | Ar | 1THGEM | 2.5×2.5 | >200 | Bondar et al, JINST 3 (2008) P07001 |
| Sheffield Univ. | Ar | 1THGEM | 4×4 | 300 | Lightfoot et al, JINST 4 (2009) P04002 |
| ETH Zurich | Ar | 1THGEM +PCB | 10×10 | 80 | A.Badertscher et al, NIM A 641 (2011) 48 |
| Budker INP | Ar | 2THGEM | 10×10 | 1000 | A. Bondar et al., in preparation |
| IRFU CEA-Saclay, ETH Zurich | Ar | 1MM | 10×10 | 5 | M. Zito et al, presented at GLA2011 (2011) |
| Budker INP | Kr | 3GEM | 2.8×2.8 | 600 | Bondar et al, NIM A 556 (2006) 273 |
| Budker INP | Xe | 3GEM | 2.8×2.8 | 200 | Bondar et al, NIM A 556 (2006) 273 |
| LIP-Coimbra | Xe | 1GEM | 2.8×2.8 | 150 | Balau et al, NIM A 598 (2009) 126 |
| Budker INP, ITEP, Weizmann Inst. | Xe | 2THGEM | 2.5×2.5 | 600 | Bondar et al, JINST 6 (2011) P07008 |

Table 1. Summary of maximum charge gains reached with MPGD multipliers operated in two-phase Ar, Kr and Xe CRADs, obtained by different groups. The charge gain is defined as that of the Budker INP and Weizmann Institute groups.

There are however several unsolved problems in two-phase CRAD performances; these are listed below.

The first problem is that the performance of MPGD multipliers in two-phase Ar and Xe is not fully understood: not all multiplier types were able to operate with electron multiplication in saturated vapour. In two-phase Ar, while G10-based THGEM multipliers successfully operated for tens of hours with gains reaching several thousands, others, namely RETHGEM and Kapton THGEMs, did not show stable multiplication in an equilibrium state [24],[75]: see Fig. 38; here gain instabilities mean either non-multiplication (with gain below 1) or large gain variations (represented by the error bars in Fig. 38). The geometrical parameters of the RETHGEMs and Kapton THGEMs were somewhat different from the regular THGEMs, namely they had smaller dielectric thickness (0.25 mm) and hole diameter (0.3 mm). In addition, in ref. [80] it was reported on the unsuccessful performance of the Micromegas multiplier in two-phase Xe: in half an hour the multiplication collapsed. The most rational explanation of these instabilities is the effect of vapour condensation within the THGEM holes or Micromegas mesh that prevents electron multiplication. The criteria for such a condensation are not yet clear. It could be that the specific properties of the holes and electrodes might play a role, i.e. the wetting capability depending on the electric-field non-uniformity and consequently on the MPGD geometry, as well as on the temperature gradients, the latter in turn depending on the electrode's heat conductivity.

The second problem is that of the gain limits of two-phase CRADs. To work with minimum threshold, the detector for rare-event experiments should be able to operate in single-electron counting mode, in a self-triggering mode and with a minimum noise level. For GEM-based two-phase CRADs this requires stable operation at gains of about 20,000 [52], while today's limit is about 4-8 times lower (see Table 1): the so far achieved level corresponds to the detection of 4 primary electrons for the triple-GEM and 20 electrons for the double-THGEM [24], at gains 5000 and 1500 respectively, i.e. it is clearly insufficient. For effective work in single-electron counting mode, the obvious way to increase the gain is an increase in the number of stages, until 4-5 stages in the case of GEMs and 3-4 stages in the case of THGEMs. In the latter case it can be technically difficult due to higher operating voltages. Therefore, it looks attractive to combine different types of MPGD multipliers, including thin and thick GEMs and Micromegas. The second way is an optical readout from THGEMs using GAPDs (see section 3.3). In this case, the THGEM gain can be reduced due to high GAPD gain.

The third problem is that of the resistance to electrical discharges of GEMs and THGEMs made of Kapton. When operating two-phase CRADs at maximum gains (approaching 10,000), it was observed that Kapton triple-GEMs were not able to withstand electrical discharges [75]: after several series of measurements the maximum reachable gain of the GEM multiplier decreased by several times. Obviously, this is due the low resistance of thin GEMs to discharges, as a result of metal evaporation from the electrodes and its deposition on the insulator in the GEM holes. The similar effect was observed for Kapton THGEMs [75]: at gains exceeding 5000 a fatal discharge occurred, resulting in irreparable damage of the THGEM (see Fig. 38). The solution of the problem might be switching to thicker THGEMs that behave in much more reliable way under discharges.

The fourth problem is that of radioactivity of G10 material of THGEMs. G10 from which THGEMs are usually made contains glass fibres, containing radioactive $^{40}$K as main source of undesirable background. To reduce background in rare-event experiments, other, radio-clean materials should be investigated: Kevlar, Kapton, etc.

Our general conclusion to this section is that the maximum gains achieved in two-phase CRADs, namely several thousands in Ar and half a thousand in Xe, though being enough for charge readout in giant liquid Ar detectors and two-phase Xe PETs (considered in section 2.4), might not be sufficient for efficient single-electron counting recording avalanche-charge in self-triggering mode, for coherent neutrino-nucleus scattering and dark matter search experiments. Accordingly, ways of increasing the overall gain should be looked for. A possible solution investigated is the optical readout of THGEM avalanches using GAPDs; it is discussed in the following section.

### 3.3 Two-phase CRADs with optical readout using GAPDs

In this section we consider a novel technique of signal recording in two-phase and gaseous CRADs. It consists of optically recording avalanche-induced scintillation light emitted from THGEM holes in Ar or Xe gas, using Geiger-mode Avalanche Photodiodes (GAPDs, [81]), thus employing the idea of combined THGEM/GAPD multiplier optical readout [26],[35],[36],[37],[38]: see Figs. 8-10 and 19.

In detectors requiring ultimate sensitivities, the optical readout using GAPDs might be preferable as compared to charge readout in terms of overall gain and noise. Indeed, the typical gains in hole-multipliers listed in Table 1, might not be sufficient for single-electron counting, recording avalanche-charge in self-triggering mode. On the other hand, even at low gas-avalanche gains, the high GAPD gain, reaching $10^6$, would substantially increase the overall gain, thus providing effective single-electron counting. In addition, multi-channel optical readout with overlapping fields-of-vision and coincidence between channels, would effectively suppress single-channel noise.

Moreover, the GAPD performance at cryogenic temperatures is superior to that at room temperature [82],[83],[84],[85],[86],[87],[88],[89]: the noise-rate is considerably reduced [82],[84],[88],[89], while the amplitude resolution [85] and the maximum gain [88] can be substantially increased. This is illustrated in Fig. 39 showing noise rate and relative Photon Detection Efficiency (PDE) characteristics. From this figure one may conclude that in two-phase Ar environment the GAPD noise rate at the PDE efficiency plateau can be done as low as few Hz only.

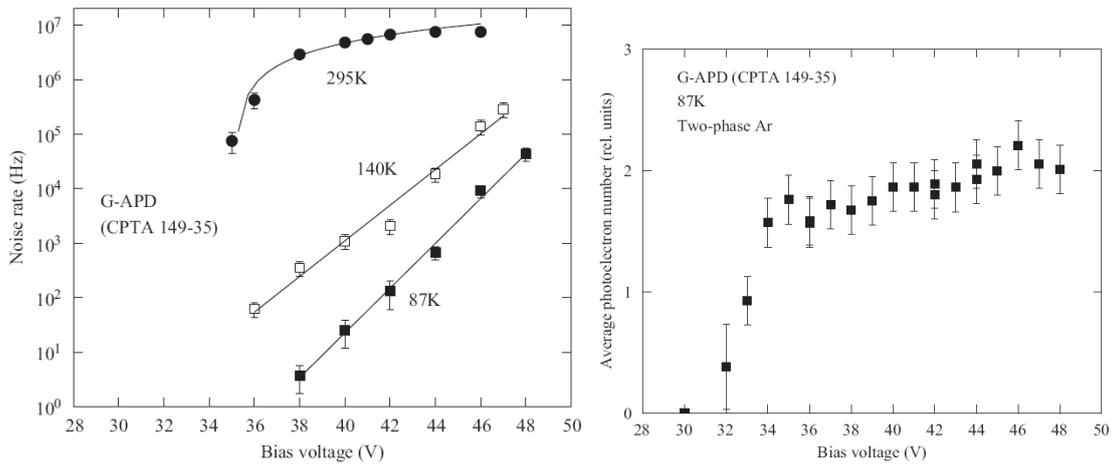

Fig. 39. Performance of the GAPD (MRS APD "CPTA 149-35" [90]) having a 2.1×2.1 mm$^2$ active area at cryogenic temperatures [88]. Shown are the GAPD noise rate (left) and relative Photon Detection Efficiency (PDE) (right) as a function of the bias voltage at different temperatures. The fits to the noise rate data points are performed with an exponential function at 87 and 140 K, and with a linear function at 295 K.

Earlier studies of optical readout from GEMs were performed in Xe at room temperature using a specific large-area APD (LAAPD) sensitive in the VUV [91],[92]. However LAAPDs have low gain (~100) that presents their main drawback when compared to GAPDs.

All noble gases have intense primary and secondary scintillations both in the VUV [15],[16] and the NIR [36],[93],[94],[95],[96],[97],[98],[99],[100],[101]: their emission spectra are shown in Fig. 40 and Figs. 41-43 respectively. Notice that in the NIR, the emission spectrum of gaseous Ar consists of atomic lines [93],[99], while that of liquid Ar is continuous [99],[100]; this will be further discussed in section 4.1. On the other hand, regular GAPDs have high quantum efficiency in the visible and NIR region, in particular of 18% on average in the region of the NIR emission of Ar, at 600-900 nm (Fig. 41). This results in two CRAD concepts with combined THGEM/GAPD multiplier optical readout: using either WLS-coated GAPDs sensitive in the VUV or uncoated GAPDs sensitive in the NIR.

As concerns the first concept, two WLS types were used to reemit the VUV light to the visible range in cryogenic environment: that of Tetraphenyl-Butadiene (TPB) [102],[103],[104],[105],[106] and that of p-terphenyl (PT) with poly-para-xylylene protective film [107],[108], the protective film being used to prevent the pollution of the noble liquid by organic WLS. Both WLS, TPB and PT, have high reemitting efficiency in the VUV, in particular in Ar and Xe emission regions, i.e. at 128 and 175 nm respectively.

The first concept was realized in the two-phase Ar CRAD using TPB-WLS coated GAPD [26] and in the two-phase Xe CRAD using PT-WLS coated GAPD matrix [38], shown in Figs. 8 and 10 respectively. Fig. 44 characterizes the performance of the two-phase Ar CRAD: THGEM light yield and THGEM charge gain characteristics are presented. Fig. 45 demonstrates the performance of the CRAD having the most sophisticated design at the moment, namely of the two-phase Xe CRAD having combined THGEM/GAPD-matrix optical readout, using Kapton THGEMs and PT-based WLS with poly-para-xylylene protective film, and PMT-matrix optical readout [38]. The figure demonstrates a typical signal of the GAPD matrix induced by avalanche scintillations in the double-THGEM multiplier. This signal is correlated to that of the PMT matrix, the latter consisting of successive signals of primary scintillations in the liquid, proportional scintillations in the gas phase in front of the first THGEM and avalanche scintillations within the THGEM holes.

The second concept was realized in the CRADs with THGEM/GAPD optical readout in the NIR in two-phase Ar [35] (shown in Fig. 9) and gaseous Xe [28]: their typical avalanche scintillation and charge signals are shown in Fig. 46. In the two-phase Ar CRAD, the time structure of the avalanche scintillation signal reflects that of the fast and slow components of the electron emission process through the liquid-gas interface (this will be discussed in section 4.3), as well as that of the ion feedback-induced avalanches [35]. The avalanche scintillation and charge signals are time and amplitude correlated. The amplitude correlation is demonstrated in Fig. 47: one can see that the combined THGEM/GAPD multiplier yields about 1 photoelectron per primary (prior to multiplication) electron, at the charge gain of 400 and at rather large GAPD viewing angle (field-of-vision angle), of ±70°. The latter could be reduced down to more practical value, namely down to ±45°, thus increasing the GAPD solid angle by a factor of 5 and allowing to effectively operate in single electron counting mode [35].

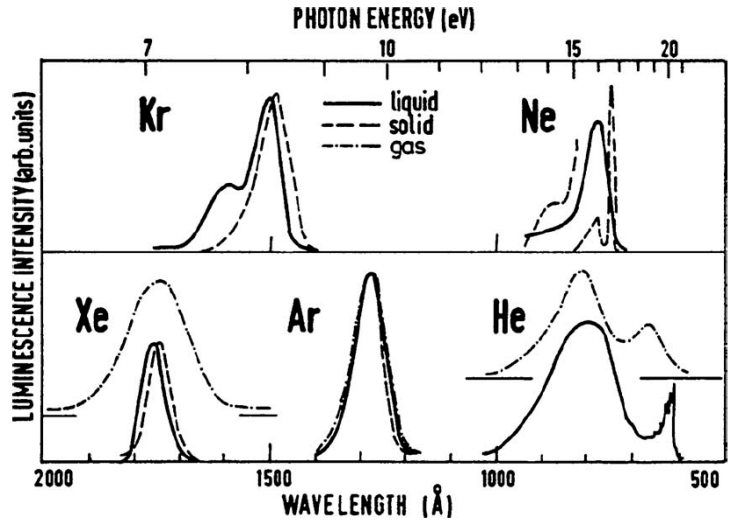

Fig. 40. Emission spectra of noble gases in the VUV, in gaseous, liquid and solid state [16].

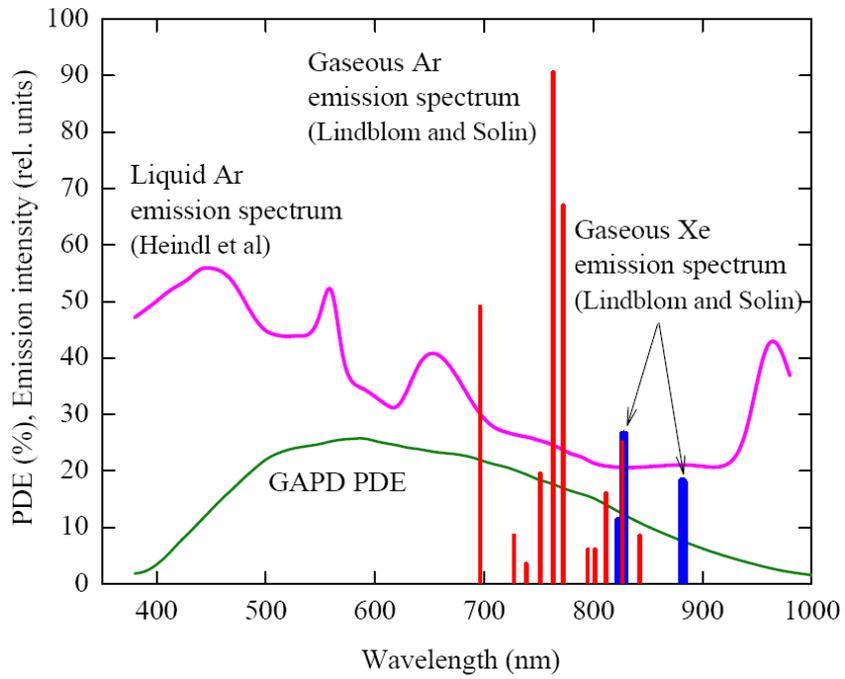

Fig. 41. Primary scintillation emission spectra of gaseous Ar [93], gaseous Xe [93] and liquid Ar [99],[100] in the visible and NIR region, in the range of 350-1000 nm, and the Photon Detection Efficiency (PDE) spectrum of the GAPD (CPTA 149-35) [90]. The relative intensity of Ar and Xe emission lines corresponds to the relative avalanche scintillation yield measured in [28].

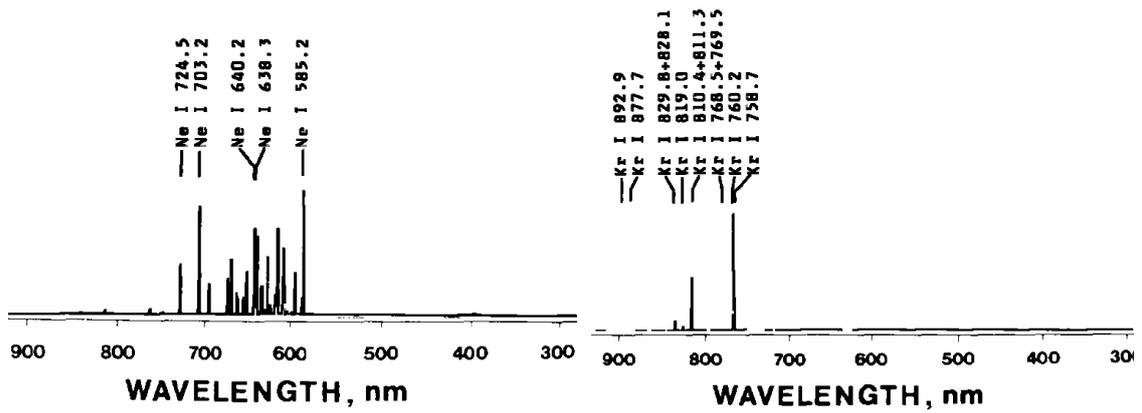

Fig. 42. Primary scintillation emission spectra of gaseous Ne (left) and Kr (right) in the visible and NIR region, in the range of 300-950 nm [93].

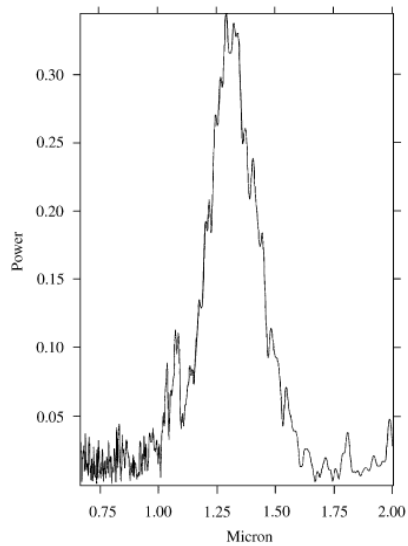

Fig. 43. Primary scintillation emission spectrum of gaseous Xe in the NIR, in the range of 700-2000 nm [96].

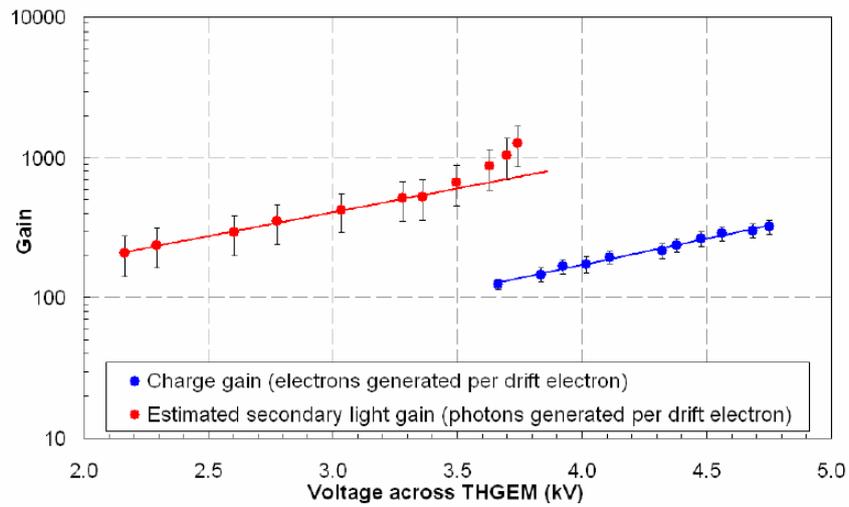

Fig. 44. THGEM light yield in the VUV (denoted as "secondary light gain") expressed in photons over 4π per primary (prior to multiplication) electron and THGEM charge gain as a function of voltage across the THGEM, in a two-phase Ar CRAD with combined THGEM/GAPD optical readout using TPB-based WLS (shown in Fig. 8) [26].

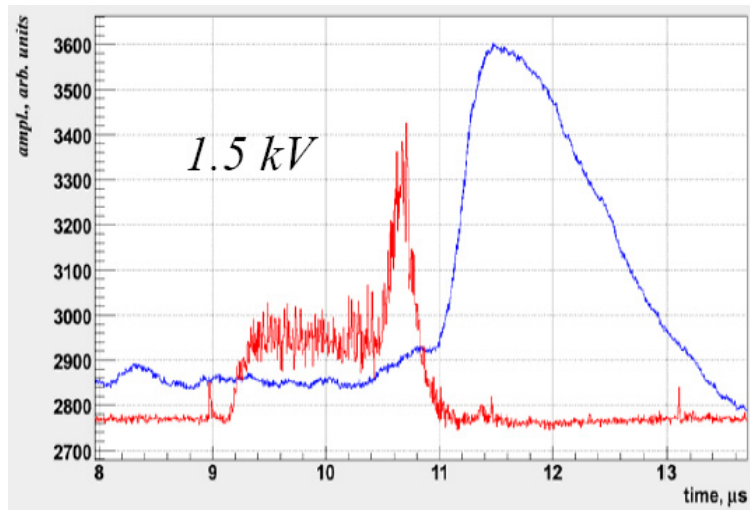

Fig. 45. Typical signals in a two-phase Xe CRAD with combined 2THGEM/GAPD-matrix optical readout in the VUV using PT-based WLS with poly-para-xylylene protective film (shown in Fig. 10) [38]. Blue trace: summed GAPD-matrix signal. Red trace: PMT-matrix signal consisting of successive signals of primary, proportional and avalanche scintillations.

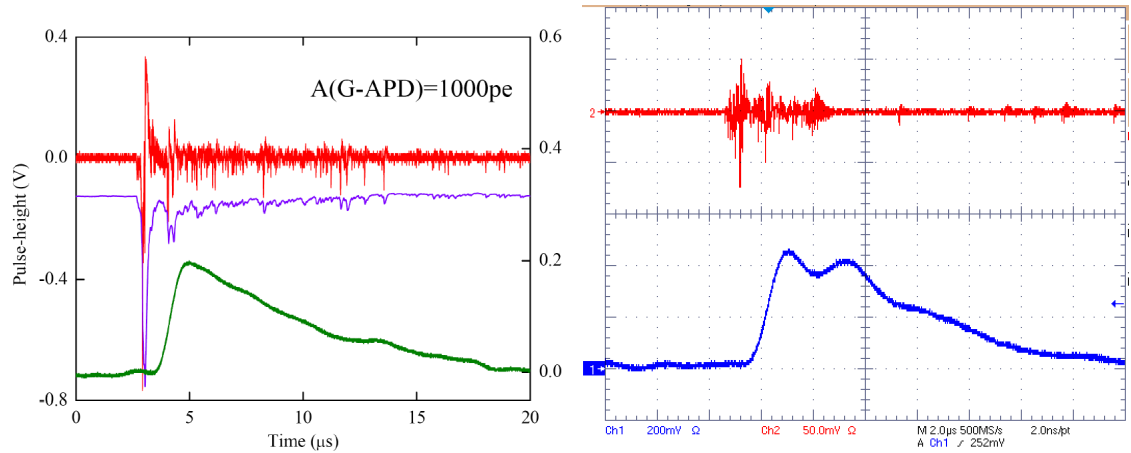

Fig. 46. Typical avalanche scintillation and charge signals in CRADs with combined THGEM/GAPD optical readout in the NIR (shown in Fig. 9), in two-phase Ar at 87 K (left) [35] and gaseous Xe at 200 K and 0.73 atm (right) [28]. GAPD avalanche scintillation signals: upper traces for bipolar pulses and middle trace for slightly filtered unipolar pulse. 2THGEM multiplier charge signals: lower traces. The signals are induced by 60 keV X-rays from $^{241}$Am source. 2THGEM charge gain: 400 (left) and 350 (right). The time scale in the right figure is 2 μs/div.

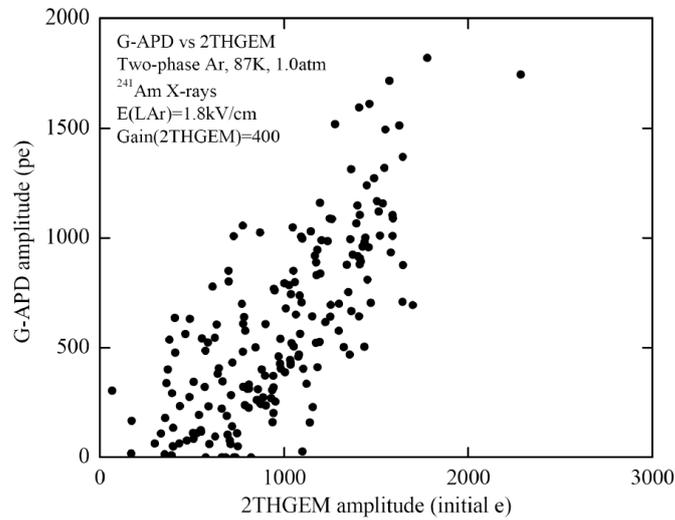

Fig. 47. Correlation between avalanche-charge and avalanche-scintillation signals in a two-phase Ar CRAD with combined THGEM/GAPD optical readout in the NIR (shown in Fig. 9) [35]. Shown is the GAPD scintillation-signal amplitude, expressed in photoelectrons not corrected for nonlinearity and cross-talk, versus 2THGEM charge-signal amplitude, expressed in primary (initial) electrons, i.e. prior to multiplication. The signals were induced by 60 keV X-rays from $^{241}$Am source. 2THGEM charge gain is 400. The GAPD viewing angle is ±70°.

The summary of combined THGEM/GAPD multiplier yields, THGEM light yields and avalanche scintillation light yields measured in two-phase Ar CRADs [26],[35] and gaseous Xe CRAD [28] is given in Tables 2 and 3 respectively. For completeness the yields obtained in gaseous Xe at room temperature [92] and in the liquid Ar CRAD [26] are also given; the R&D results for the latter will be discussed in section 4.2.

In these tables, the combined THGEM/GAPD multiplier yield, expressed in photoelectrons per primary (prior to multiplication) electron, characterizes the overall efficiency of optical readout for a given CRAD design: it depends on the charge gain, GAPD solid angle, noble gas emission intensity and GAPD PDE spectrum. In two-phase Ar CRADs, this yield was measured to be about 1 photoelectron per primary electron, both for readout in the VUV and the NIR (see Table 2). This clearly indicates upon reaching the single electron counting sensitivity in two-phase Ar CRADs with THGEM/GAPD optical readout at moderate charge gains, in excess of several hundreds, successfully demonstrating the proof-of-principle of the concept.

On the other hand, the THGEM/GAPD multiplier yield in Xe in the NIR is about an order of magnitude lower than that in Ar (see Table 3); it is in accordance to the overlap of the GAPD's PDE spectrum with that of the noble-gas emission in the NIR range (limited to 950 nm) (Fig. 41). Due to these facts, Xe-based detectors with optical readout using uncoated GAPDs should be considered inapplicable in rare-event experiments requiring single-electron sensitivity, in contrast to Ar-based detectors. At the same time, the expected THGEM/GAPD yield, of the order of 1000 photoelectrons per 511 keV $\gamma$-ray as deduced from table 3, is sufficient for PET applications. In addition, with Xe, better results could be expected with WLS-coated GAPDs, sensitive to its more copious VUV emission (see next paragraph). An alternative solution for THGEM optical readout in Xe in the NIR might be InGaAs photodiodes of high sensitivity up to 1700 nm [95], i.e. within the major Xe emission range [96] (see Fig. 43).

The avalanche-scintillation light yield over $4\pi$, expressed in photons per avalanche electron, characterizes the noble gas emission intensity in a given spectral range, VUV or NIR. It should be remarked that these light yields, obtained in the THGEM in two-phase Ar CRADs in the VUV [26] and NIR [35], and in the GEM in gaseous Xe in the VUV [92], of about 8, 4 and 3 photon per avalanche electron respectively (see Tables 2 and 3), turned out to be rather high. These should be compared to that of ~1 NIR photon per avalanche electron presented in [94] for avalanche scintillations in Ar in a parallel-plate chamber. Such a difference in light yields could be explained by strong dependence on the gas amplification structure, observed in particular in [94]. That means that hole-type amplifying structures, namely GEMs, THGEMs and MHSPs, might be particularly efficient in terms of avalanche scintillations.

| Medium conditions for secondary scintillations recorded | Gaseous Ar at 87K and 1.0atm in two-phase mode | Gaseous Ar at 87K and 1.0atm in two-phase mode | Liquid Ar at 87K |
|---|---|---|---|
| Combined multiplier type | 1THGEM/WLS/GAPD | 2THGEM/GAPD | 1THGEM/WLS/GAPD |
| THGEM multiplier charge gain | ~120 | 400 | Unknown |
| THGEM/GAPD yield, photoelectrons per primary electron (prior to multiplication) | ~1.4 | 0.7 at GAPD viewing angle of ±70° | |
| THGEM light yield over 4π, photons per primary electron | ~900 photons in VUV | 1500 photons in NIR | ≤500 photons in VUV |
| Avalanche scintillation light yield over 4π, photons per avalanche electron | 8 photons in VUV | 4 photons in NIR | |
| Reference | [26] | [35] | [26] |

Table 2. Summary of combined THGEM/GAPD multiplier yields, THGEM light yields and avalanche scintillation light yields, measured in two-phase Ar CRADs [26],[35] and in liquid Ar CRAD [26].

| Medium conditions for secondary scintillations recorded | Gaseous Xe at 200K and 0.73atm | Gaseous Xe at room T and 1.5atm |
|---|---|---|
| Combined multiplier type | 2THGEM/GAPD | 1GEM/LAAPD |
| Charge gain of THGEM or GEM multiplier | 350 | 900 |
| THGEM/GAPD yield, photoelectrons per primary electron (prior to multiplication) | 0.07 at GAPD viewing angle of ±70° | |
| THGEM or GEM light yield over 4π, photons per primary electron | 240 photons in NIR | 3000 photons in VUV |
| Avalanche scintillation light yield over 4π, photons per avalanche electron | 0.7 photons in NIR | 3 photons in VUV |
| Reference | [28] | [92] |

Table 3. Summary of combined THGEM/GAPD and GEM/LAAPD multiplier yields, THGEM and GEM light yields and avalanche scintillation light yields, measured in gaseous Xe at cryogenic temperature [28] and at room temperature [92].

## 3.4 CRADs as cryogenic GPMs

In this section Gaseous Photomultipliers using solid photocathodes (GPMs; see for example reviews [109],[110]) operated at cryogenic temperatures are discussed. There are two types of cryogenic GPMs: those operating in a sealed mode, i.e. with window separating the GPM from the detection medium [29],[30],[31],[54] (Figs. 11 and 12) and those operating in a windowless mode [32],[44],[45],[54] (Fig. 6), i.e. directly in the noble gas detection medium. Cryogenic GPMs of the second type, with CsI photocathode operated in dense pure noble gases, turned out to have poor efficiency [32],[44],[54], presumably due to strong photoelectron backscattering and the rather low photoelectron collection efficiency. This was discussed in particular in section 3.2 in the case of the two-phase Ar CRAD with GEM multiplier readout and CsI photocathode on the first GEM [32].

Accordingly in this section, we focus upon cryogenic GPMs of the first type: their main advantage consists in using the noncondensable gas mixture with molecular additives, permitting to attain high gains at cryogenic temperatures. In particular we consider here the operation results for the GPM used in the liquid Xe CRAD with cryogenic GPM separated by $MgF_2$ window from the noble liquid [30],[31]: it is shown in Fig. 12. The GPM consisted of a reflective CsI photocathode deposited on top of a THGEM; further multiplication stages were either a second THGEM or a Parallel Ionization Multiplier (PIM) followed by a Micromegas (MM).

Gains of $10^4$ were measured with a CsI-coated double-THGEM multiplier in $Ne/CH_4$ and $Ne/CF_4$ mixtures at 173 K [30]. Even higher gains, exceeding $10^6$, were attained in a triple-structure THGEM/PIM/MM in $Ne/CF_4$ mixture at 171 K [31]: see Fig. 48. Scintillation signals induced by alpha particles in liquid Xe were successfully measured there with a double-THGEM cryogenic GPM in $He/CH_4$ and a triple-structure cryogenic GPM in $Ne/CH_4$ [30]: see Fig. 49.

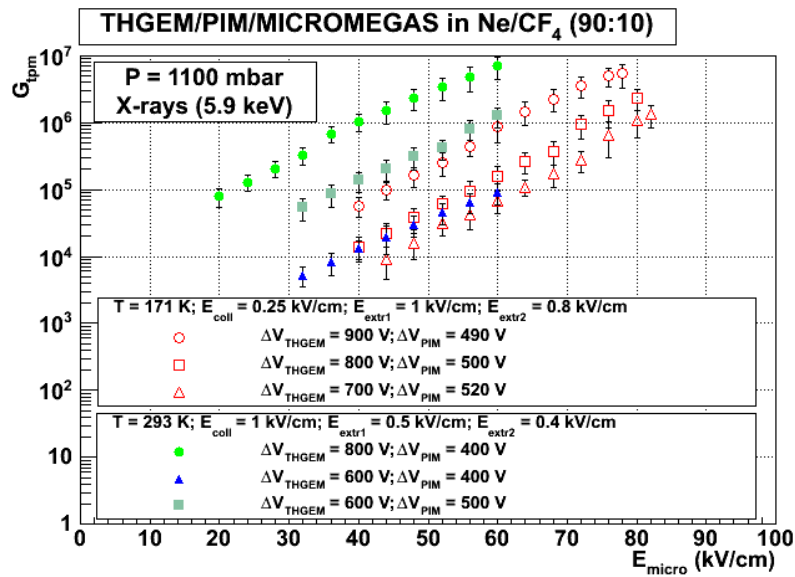

Fig. 48. Charge-gain curves obtained with the THGEM/PIM/MM multiplier at 293 K and 171 K in 1100 mbar of Ne/CF4 (90:10) mixture for different THGEM and PIM voltages [31].

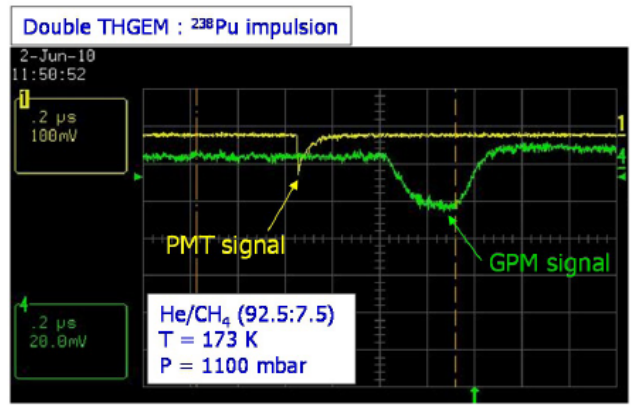

Fig. 49. Scintillation signals induced by $^{238}$Pu α-source recorded in a liquid Xe CRAD with cryogenic GPM separated by MgF$_2$ window from the noble liquid, in coincidence between the GPM (double-THGEM) and the PMT [30]. Gas mixture: He/CH4 (92.5:7.5), T = 173 K and P = 1100 mbar.

## 4. Selected CRAD physics effects

In this chapter we discuss the most remarkable physical effects governing the performance of gaseous, two-phase and liquid CRADs. Some of these effects remain unconfirmed and not fully understood. Accordingly, further studies are needed in these directions.

### 4.1 NIR scintillations in noble gases and liquids

The study of NIR scintillations in noble gases and liquids has been triggered by the results on the two-phase Ar CRAD performance with combined THGEM/GAPD optical readout in the NIR, discussed in section 3.3: a rather high avalanche (secondary) scintillation yield in the NIR, of about 4 photons per avalanche electron, was reported there [35].

Until recently, noble gas scintillations in high energy physics experiments have been recorded essentially in the VUV, necessitating the use of sophisticated VUV-sensitive photodetectors. Indeed, in the VUV the primary scintillation yield is rather high: of about (40-60)×$10^3$ photon/MeV in liquid Ar and Xe [15],[16] and 14×$10^3$ photon/MeV in gaseous Xe [111]. The VUV emission is caused by reactions between excited and ionized atoms producing excimers which decay radiating the VUV continua [112]: see Fig. 40 showing the appropriate emission spectra in all noble gases. At high pressures this emission was generally believed to dominate over all other types of radiative decays such as atomic emission in the visible and infrared regions [15],[16].

On the other hand, as early as 20 years ago it was suggested that this statement might not be valid due to the discovery of intense atomic emission scintillations in practically all noble gases in the near infrared (NIR) [93]: see Figs. 41-43 showing the appropriate emission spectra in Ne, Kr, Ar and Xe. In particular in Ar this spectrum is in the wavelength range of 690-850 nm [93] (Fig. 41) and in Xe at 800-1600 nm [96] (Fig. 43). Since then, the NIR emission spectra of scintillations in gaseous and liquid Ar and Xe have been further studied in several works: both for primary [96],[97],[99],[100] and secondary [94],[95] scintillations. In gaseous Ar this kind of scintillation was attributed to transitions between the atomic states of the Ar (3p$^5$ 4p) and Ar (3p$^5$ 4s) configurations [93],[94],[100]. In contrast, the emission spectrum of liquid Ar in the NIR is continuous [99],[100] (see Fig. 41); its emission mechanism has not been yet clarified. However, little was known about the absolute NIR scintillation yield in noble gases: almost nothing about that in Ar and only the lower limit in gaseous Xe ($\geq 21 \times 10^3$ photon/MeV) [95],[98].

In this section we present the experimental data obtained recently in ref. [36], in support for the hypothesis of intense NIR scintillations in Ar in view of its potential application in rare-event experiments: the scintillation yields in gaseous and liquid Ar were measured in the NIR and visible region using GAPDs. The latter have rather high PDE in the wavelength range of 400-1000 nm, of about 15% on average [90], providing direct and effective detection of NIR scintillations without WLS.

It was confirmed that in gaseous Ar at cryogenic temperatures, the non-VUV scintillations took place essentially in the NIR [36]: the primary scintillation yield was measured to be comparable to that in the VUV, amounting to 17000 ± 3000 photon/MeV in the range of 690-1000 nm, at temperatures of 163 and 87 K. This is seen from Fig. 50 where the data points obtained at lower electric fields, below 1 kV/cm, are due to primary scintillations.

At higher electric fields, the scintillation yield increases with field due to secondary proportional scintillations, i.e. due to electroluminescence. Here the notion "electroluminescence", defined in its narrow sense, is equivalent to that of "proportional scintillations", the scintillation intensity of the latter being proportional to the electric field. Proportional scintillations in noble gases are caused by atomic excitation processes at moderate electric fields. At higher fields these are taken over by atomic ionization processes, i.e. by those of electron avalanching. Accordingly, at higher fields proportional scintillations are taken over by avalanche scintillations, the intensity of the latter being not proportional to the electric field.

The reduced electroluminescence yield is defined as $Y_{el}/N$, where N is the atomic density and $Y_{el}$ is the electroluminescence yield. In a parallel plate gap, the electroluminescence yield is defined as the number of photons ($N_{ph}$) normalized to the total ionization charge generated in the gap ($N_e$), the primary scintillation contribution being subtracted, and to the average electron drift path in the gap (d):

$$Y_{el} = N_{ph}/N_e/d \ .$$

The reduced electroluminescence yield is shown in Fig. 51: its universally valid amplification parameter at 163 K (the slope of the line in Fig. 51) was measured to be 13 photons per drifting electrons per kV. It should be remarked that recent simulations of the Ar electroluminescence yield in the NIR were in fair agreement with the experimental data [101][113]. Finally, though having somewhat lower yield than that in the VUV (shown in Fig. 51 for comparison [114]), proportional scintillations may substantially increase the scintillation yield in the NIR as compared to that of primary scintillations: by an order of magnitude, to hundreds of thousands photons per MeV (Fig. 50).

In liquid Ar, the primary scintillation yield was measured to be considerably reduced compared to that of gaseous Ar, amounting to 510 ± 90 photon/MeV in the range of 400-1000 nm (see Fig. 50). It should be remarked that in contrast to gaseous Ar, no secondary scintillations have been observed in liquid Ar up to the electric fields of 30 kV/cm.

There is potentially a wide variety of applications of noble gas NIR scintillations in high energy physics experiments. Among them are two-phase Ar detectors with THGEM/GAPD-matrix optical readout in the NIR for coherent neutrino-nucleus scattering and dark matter search experiments considered in section 2.4 [36]: see Fig. 19. In section 3.3 it was demonstrated that such a detector can operate in single electron counting mode at charge gains exceeding 400 [35]. A practical detector of this type would comprise GAPDs matrices placed a few millimeters behind THGEM multipliers, with a pitch of ~1 cm, viewing clusters of multiplier holes under an angle of ±45°; these would cover the detector's active area with spatial resolutions sufficient for rare-event experiments. For example, for a 100 kg liquid Ar TPC of a volume of 40×40×40 cm$^3$ the total number of GAPDs would be reasonable, of about 1600. Such a detector would be robust, stable, simple and relatively cheap.

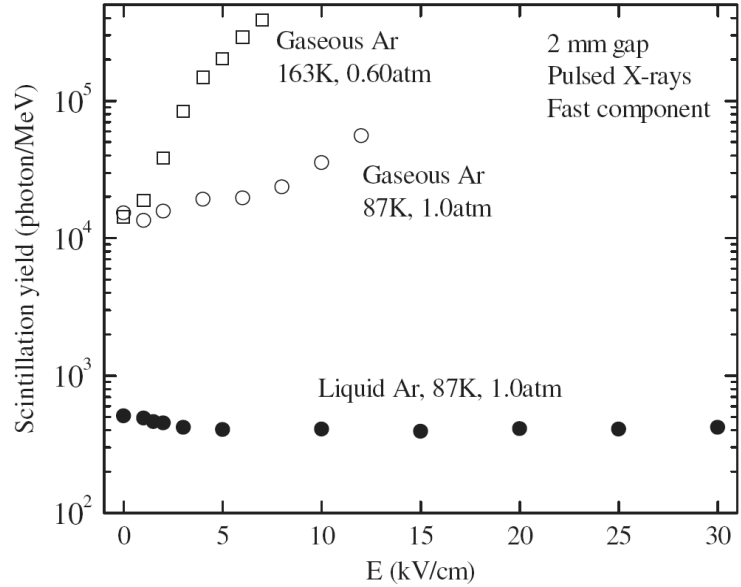

Fig. 50. Scintillation yield in liquid Ar in the NIR and visible region, in the range of 400-1000 nm, as a function of the electric field, at 87 K and 1.0 atm, and that in gaseous Ar in the NIR, in the range of 690-1000 nm, at 163 K and 0.60 atm and at 87 K and 1.0 atm [36]. The scintillation yield is given in the number of photons per MeV of deposited energy of the primary ionization. The data were obtained under X-ray irradiation in a 2 mm thick gap.

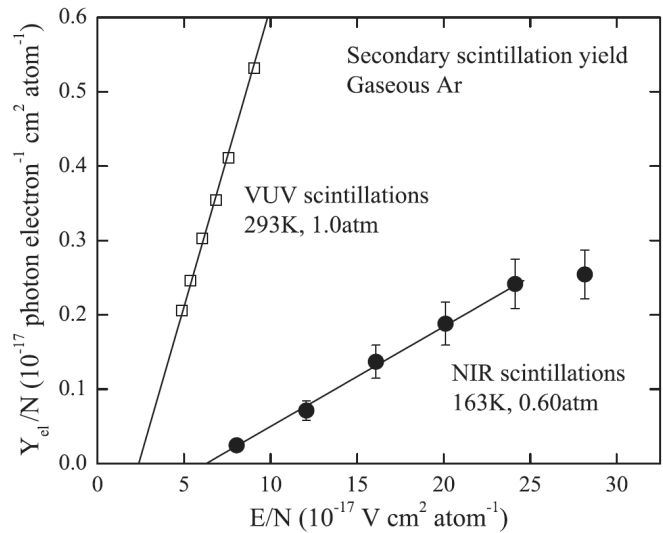

Fig. 51. Reduced electroluminescence yield in gaseous Ar in the NIR, in the range of 690-1000 nm, as a function of the reduced electric field, measured at 163 K and 0.60 atm [36]. For comparison that of the VUV measured at room temperature [114] is shown.

The other application of NIR scintillations might be noble liquid non-VUV scintillation calorimetry. The scintillation yield in liquid Ar measured here in the NIR and visible region (~500 photon/MeV) might be enough for high energy calorimetry: it is comparable with that of the fast solid scintillators being already used in calorimeters, namely higher than that of PbWO$_4$ (~100 photon/MeV) and somewhat lower than that of pure CsI (~2000 photon/MeV). The NIR scintillation yield in liquid Xe is expected to be of the same order as that of Ar, providing the applicability in liquid Xe NIR scintillation calorimetry. The readout of the calorimeters might be performed using NIR-sensitive photodetectors: Si APDs and GAPDs for liquid Ar and InGaAs photodiodes for liquid Xe.

The next possible application field of NIR scintillations in noble gases is the imaging technique using either Si-based CCDs for recording Ne, He and Ar emission or InGaAs CCDs for recording Kr and Xe emission. Indeed, CCDs have rather high sensitivity both in the visible and NIR regions [115]: in particular the high quantum efficiency of Si-based CCDs, of about 30% at 800 nm and 70% at 500 nm, matches very well emission spectra of gaseous Ar and Ne and liquid Ar respectively (see Figs. 41 and 42). One such application is the project for directional solar neutrino detection using Ne and He gaseous or two-phase detectors with combined GEM/CCD optical readout considered in section 2.4 [33],[34]: see their concept in Fig. 20 and proof-of-principle in Fig. 28. The other application is the two-phase detector with GEM/CCD or CCD optical readout for digital radiography considered in section 2.4 (see Fig. 22).

## 4.2 VUV electroluminescence in liquid Ar

In this section we discuss the results on the VUV electroluminescence in liquid Ar obtained in ref. [26], in the experimental setup shown in Fig. 8: the successful operation of the liquid Ar CRAD with GAPD optical readout (using WLS) of a THGEM plate immersed in the liquid was demonstrated. The appropriate concept for such a CRAD is depicted in Fig. 13 [55],[56]. The concept is based on the physical effect of noble liquid electroluminescence. Until recently the only evidence for such an effect was presented in ref. [116]: liquid Xe electroluminescence was observed using thin wires at very high electric fields, of the order of 0.5 MV/cm.

The mystery of the results obtained in [26] is that liquid Ar electroluminescence took place at much lower fields: this is seen from Fig. 52 showing the THGEM light yield in liquid Ar as a function of the electric field in the center of the THGEM holes. We can speak here about the electroluminescence effect, i.e. about that of proportional scintillations, since their intensity is proportional to the electric field. Electroluminescence starts at a field of 58 kV/cm, reaching a rather high yield at 63 kV/cm, of about 500 photon in the VUV over 4π per primary electron. These electric fields are far less than those of 3 MV/cm expected from theoretical calculations for electroluminescence in liquid Ar due to atomic excitation mechanisms [57]. It is also surprising that the THGEM light yield obtained in liquid Ar is of the same order as the THGEM light yield obtained in gaseous Ar at a charge gain of about 100 (operated in the two-phase Ar CRAD): see Table 2.

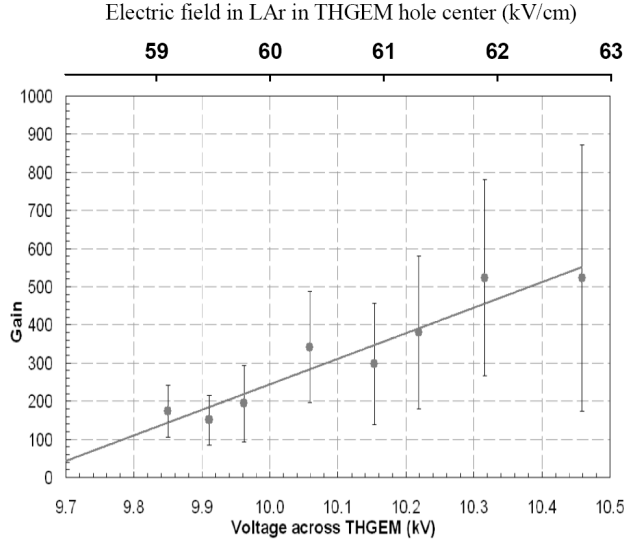

Fig. 52. THGEM light yield in liquid Ar (denoted here as "gain"), expressed in VUV photons over 4π per primary electron, as a function of the voltage across the THGEM, in a liquid Ar CRAD with combined THGEM/GAPD optical readout using WLS (shown in Fig. 8) [26]. The appropriate electric field in the THGEM hole center is shown on the top axis.

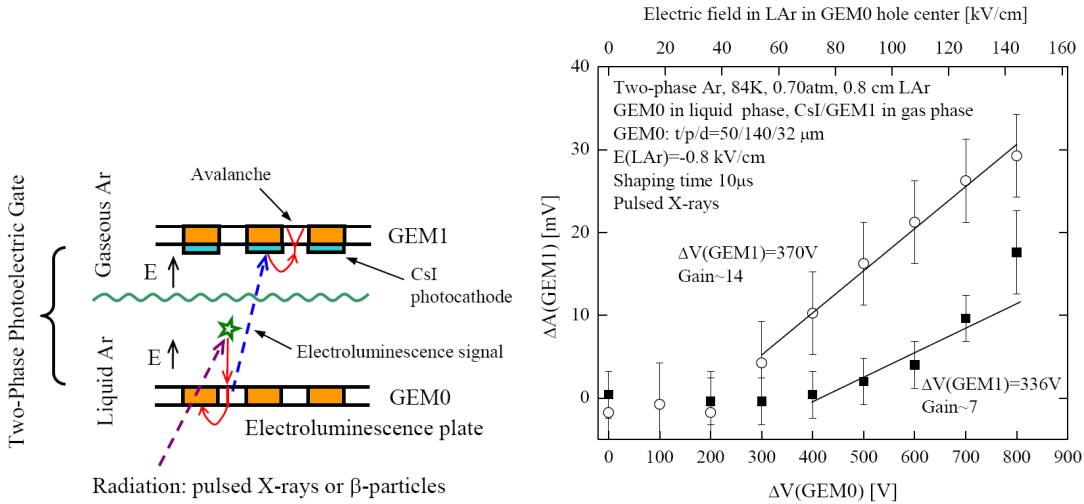

Fig. 53. Left: schematics of the experimental setup to study the liquid Ar VUV electroluminescence in a GEM using the concept of Two-Phase Photoelectric Gate [59]. Right: amplitude of the electroluminescence signal generated in a GEM plate immersed in liquid Ar (GEM0), measured with a cryogenic GPM with CsI photocathode (GEM1), as a function of the voltage across the GEM0 [59]. The appropriate electric field in the GEM0 hole center is shown on the top axis.

It is interesting that there was an indication on the similar electroluminescence effect in liquid Ar [59]; the effect was observed in the course of the study of the two-phase Ar CRAD with Two-Phase Photoelectric Gate [58] considered in section 2.3 (Fig. 14). To conduct this study the experimental setup shown in Fig. 53 (left) was used [59]. The VUV electroluminescence signal, generated in a GEM plate immersed in liquid Ar, was observed using a windowless cryogenic GPM with CsI photocathode based on the GEM multiplier: see Fig. 53 (right). It is indicative that the electric fields in the center of the GEM holes in liquid Ar, at which the electroluminescence signal started to be observed (Fig. 53), of about 50-60 kV/cm, were very similar to those of ref. [26] (Fig. 52). This correspondence can hardly be accidental.

At the moment the effect of liquid Ar electroluminescence in THGEM and GEM plates is not understood. One hypothesis is that of the role of uncontrolled impurities in the liquid: the impurities might somehow produce electroluminescence at a threshold much lower than that expected in pure liquid Ar. The other explanation might the presence for some reason of gaseous bubbles associated to the THGEM and GEM holes, within which the electron avalanches could develop producing avalanche scintillations. The latter hypothesis can be confidently tested by measuring the electroluminescence emission spectrum in the NIR: the hypothesis would be confirmed if the emission spectrum would consist of atomic lines corresponding to gaseous Ar (see Fig. 41 and discussion in section 4.1). Otherwise, the continuous NIR spectrum would indicate on the real liquid Ar emission.

## 4.3 Electron emission in two-phase CRADs

One of the most important physical effects governing the performance of two-phase CRADs is that of electron emission through the liquid-gas interface [78],[117],[118],[119]. In two-phase Ar such a process has fast and slow electron emission components, lasting for less than a nanosecond and over few microseconds respectively [78],[117],[118]. The fast component was explained by emission of "hot" electrons heated by an electric field when drifting in the liquid and having overcome a potential barrier at the liquid-gas interface [14],[17],[120]: see Fig. 54 showing the potential energy diagram and physical processes in two-phase system [78]. The slow component was explained by thermionic emission of "cold" electrons, including those cooled down after reflections from the potential barrier [118],[120] and perhaps after photoelectron backscattering from the molecules in the gas phase [78].

It should be emphasized that the slow component has never been observed in two-phase Kr and Xe systems [14],[17],[120], presumably due to higher potential barrier as compared to Ar. And vice versa, the fast component has never been observed in two-phase He and Ne systems [119],[120], since the electrons in liquid He and Ne are localized in the bubbles and thus cannot be heated by the electric field. Consequently, the two-phase Ar systems are unique in terms of providing the opportunity to study directly both fast and slow electron emission processes.

Such a study has been recently carried out in two-phase CRADs operated in Ar and Ar+$N_2$ with GEM multiplier charge readout [78]; here the fast response of the GEM multiplier provided the required time resolution. In the Ar+$N_2$ system, the $N_2$ content was 0.5% in the liquid and 1.5% in the gas phase. In both systems, the fast and slow components of the electron emission through the liquid-gas interface have been observed directly: see Fig. 55. In Ar, the slow emission component dominated even at higher electric fields, reaching 2 kV/cm: see Fig. 55 (left). On the contrary in Ar+$N_2$, the interesting physical effect was observed: the fast emission component dominated, the slow component being almost fully converted to the fast component at fields exceeding 1.5 kV/cm. This is seen in Fig. 55 (right). Such behaviour was explained by suppression of the electron backscattering effect in the gas phase of the Ar+$N_2$ system. In other words, the slow electron emission component in two-phase systems was supposed to appear whenever the potential barrier is low and the backscattering effect is strong, in particular in two-phase Ar systems.

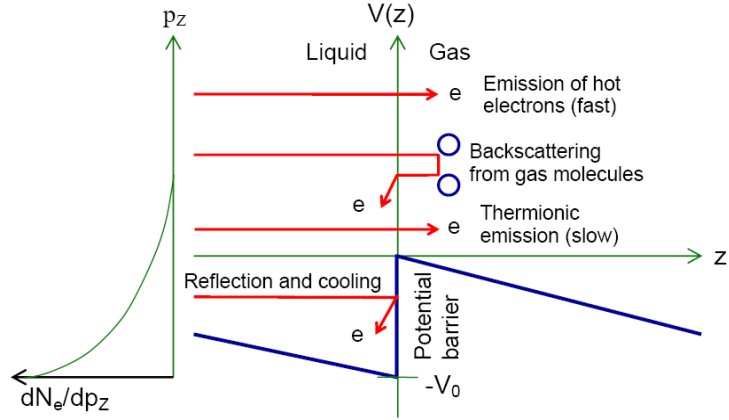

Fig. 54. Potential energy diagram and physical processes at the liquid-gas interface in two-phase (liquid-gas) system [78].

In addition, the electron emission efficiency in two-phase Ar+$N_2$ was found to be similar to that of Ar; its value was approaching to 100% even at lower electric fields. This is totally different from two-phase Kr and Xe systems, where the electron emission efficiency has specific threshold behaviour as a function of the field [14],[17].

Our general conclusion is that two-phase Ar+$N_2$ CRADs might be superior to those of other noble gases. They may have fast signals due to intrinsically fast avalanche signals [11], the slow electron emission component being suppressed [78]. Also they may have higher avalanche gains in the gas phase compared to pure Ar [11],[48]. Scintillation and ionization detection properties in liquid Ar doped with $N_2$ were studied in refs. [121],[122],[123], from which one can learn that VUV scintillations in liquid Ar+$N_2$ are suppressed. Therefore such detectors might be relevant to those experiments where mainly the ionization signal is recorded, in particular to coherent neutrino-nucleus scattering experiments and large-scale neutrino detectors.

In the rest of the section we discuss the unexpected physical effect on charge transmission through the liquid Ne surface observed more recently in ref. [119]. The motivation for that study stemmed from the E-bubble project for solar neutrino detection based on the two-phase He and Ne detectors with GEM/CCD optical readout [33] considered in section 2.4 (see Fig. 20). It was observed that in Ne (contrary to He), instead of expected smooth electron emission through the liquid-gas interface from electron bubbles [124],[125], the periodic charge eruption occurred, ejecting droplets of Ne which disrupt the surface after eruptions [119]: see Fig. 56. This leads one to conclude that the effective trapping times of the electrons under the liquid surface are longer than the trapping times reported earlier [125]: of a few tens of seconds. Accordingly, the presented phenomena eliminate the controlled transport of signal charges from the liquid phase into the gas phase in Ne. Just this conclusion prompted the modification of the E-bubble project concept [33], suggesting the high-pressure Ne CRAD instead of that of two-phase [34].

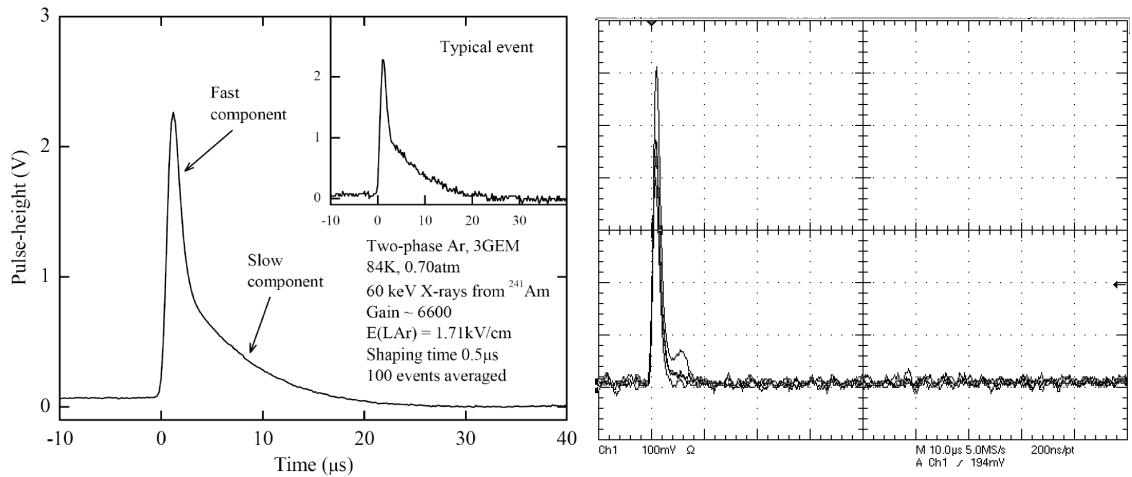

Fig. 55. Typical anode signals in two-phase CRADs with GEM multiplier charge readout at an electric field within the liquid of about 1.7 kV/cm [78]. Left: in Ar; the fast and slow components are distinctly seen, the slow component being dominated. Right: in Ar+$N_2$ (0.5% in the liquid, 1.5% in the gas); the fast component is mostly seen, the slow component being almost fully converted to the fast component. The time scale in both figures is 10 μs per division.

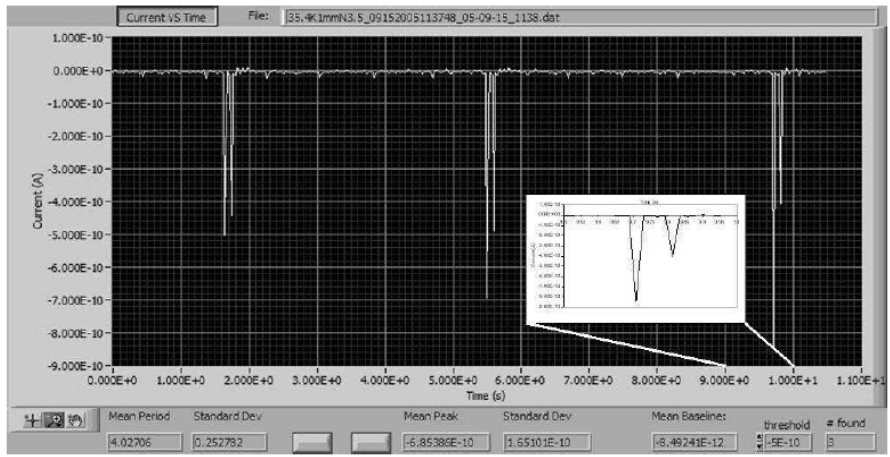

Fig. 56. Periodic current pulses due to charge eruption through the liquid-gas interface in two-phase Ne at 27 K [119]. The primary ionization is produced in liquid Ne by α-particles. The inset shows a zoom of the time span from 9-10 s where the multiple spikes in the current trace are clearly visible.

## 4.4 Electron avalanches at low temperatures

In this section we discuss electron avalanche mechanisms in dense noble gases at low temperatures. Understanding such mechanisms is of primary importance for CRAD performances. Little is known however about the physics of electron avalanching in noble gases at low temperatures. To the best of our knowledge, there are only a few works dealing with this matter: in proportional multiplication mode - in dense Ar [6],[19] and Kr [19] at around 120 K, in low-dense Ar at 77 K [126], in low-dense He at 4.2 K [7] and in dense He and Ne down to 2.6 K [20],[21]; in glow discharge mode - in He near and below 77 K [127] and in dense Ar near 150 K [128].

As concerns electron avalanching in dense Ar and Kr at cryogenic temperatures, in a proportional mode, it almost does not differ from that at room temperature [19], i.e. it is described by the standard mechanism of electron impact ionization [18]. This statement is expected to be valid also for Xe [28].

Regarding the glow discharge at low temperatures, it was supposed to be governed by new mechanisms, other than that of electron impact ionization, namely in He by the Penning-like mechanism which operates through the long-lived metastable excited atoms [127]:

$$He^m + He^m \rightarrow He + He^+ + e\,;$$

and in Ar by the associative ionization mechanism which operates through the short-lived resonance excited states [128]:

$$Ar^* + Ar \rightarrow Ar_2^+ + e\,.$$

It is interesting that the associative ionization was initially supposed to be responsible also for the high gain operation of GEMs in dense He and Ne at room temperature [18]; however it was presumably the erroneous hypothesis. Indeed, it was shown [20],[21] that electron avalanching in proportional mode in dense pure He and Ne is described by the standard mechanism of electron impact ionization even at very low temperatures, in particular down to 2.6 K in He. This is seen from Fig. 57 showing a comparison of ionization coefficients in dense He at low temperatures obtained from single-GEM gain-voltage characteristics [20], with those taken from the literature [129] obtained at room temperature and low densities. One can see that the data at temperatures below 20 K are in good agreement with those at room temperature, i.e. these are well described by the electron impact ionization mechanism.

On the other hand, the ionization coefficients at temperatures above 60 K obtained in [20] were significantly enhanced (see Fig. 57), resulting in high GEM gains observed in He and Ne above 77 K. As discussed in section 3.1, these high gains were due to the Penning effect in uncontrolled ($\geq 10^{-5}$) impurities, most probably in the $N_2$ impurity:

$$He^m + N_2 \rightarrow He + N_2^+ + e\,.$$

At lower temperatures these impurities froze out, resulting in the considerable gain drop observed at temperatures below 40 K [20],[21]: see Fig. 27 (left).

Long-lived metastable states play a crucial role in this Penning mechanism. Evidence for their presence in an avalanche was obtained by analyzing the time structure of the avalanche signals. It was observed that in He at 62 K and in Ne at room temperature, the avalanche development in time in the triple-GEM multiplier was rather slow: the avalanche signal had an unexpectedly large delay with respect to the primary ionization signal. The delay was of the order of 10 μs: this is seen from Fig. 58. In addition, the avalanche delay turned out to be a logarithmic function of the gain. This dependence was explained in the frame of a simple model of the avalanche development [20]:

$$T = \tau(\ln G - 2\ln\varepsilon).$$

Here, $T$ is the avalanche delay, $\tau$ the life-time of the excited atom, $G$ the triple-GEM gain, $\varepsilon$ the charge transfer efficiency from GEM output to the following elements. Thus, the line slope in Fig. 58 provides an estimation of the life-time of an excited atom: $\tau = 2.4$ μs. It is apparent that only metastable atoms can live such a long time, thus confirming the Penning mechanism of the avalanche development at temperatures above 40 K.

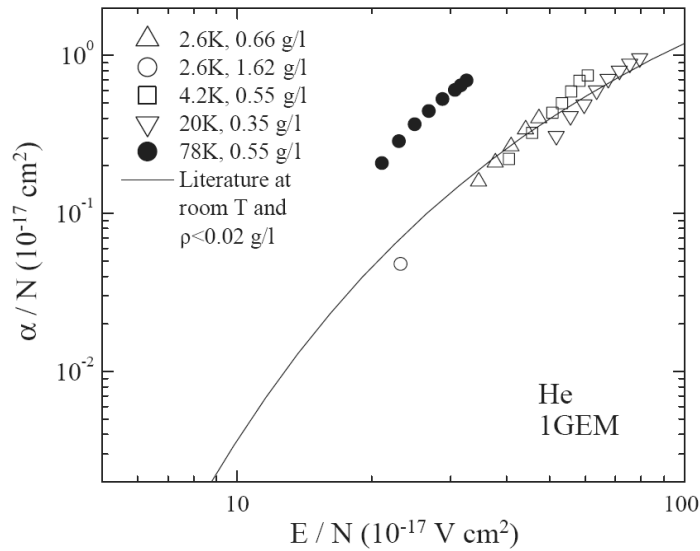

Fig. 57. Reduced ionization coefficients as a function of the reduced electric field in dense He at low temperatures, obtained from single-GEM gain-voltage characteristics [20]. The data are compared to those taken from literature [129], obtained at room temperature and low densities.

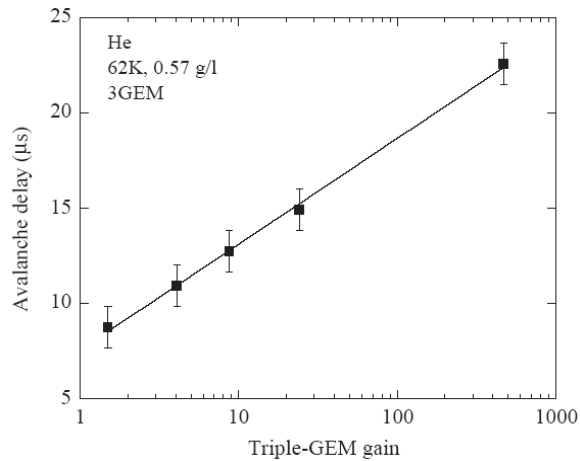

Fig. 58. Delay of the avalanche signal in the triple-GEM multiplier with respect to the primary ionization signal as a function of the gain in He at 62 K [20], i.e. most probably in the Penning mixture of He and $N_2$ impurity with the impurity content of the order of $5\times10^{-5}$.

## 5. Conclusions

Originally given rise from the fact of the high-gain operation of GEMs in pure noble gases, the idea of Cryogenic Avalanche Detectors (CRADs) had triggered intense and difficult R&D work in the course of last 8 years. This resulted in a variety of advanced CRAD concepts developed in this period. For the time being the most promising and intensively studied concepts are those of two-phase CRADs with THGEM multiplier charge readout, optical readout of CRADs with combined THGEM/GAPD multipliers and CRADs with cryogenic GPMs based on MPGDs.

Such kinds of CRADs may come to be in great demand in rare-event experiments, such as those of coherent neutrino-nucleus scattering, dark matter search and giant liquid Ar detectors for (astrophysical) neutrino physics, as well as in medical imaging fields, such as PET and digital radiography.

In addition, this R&D work has significantly advanced the understanding of a number of remarkable physical effects related to CRAD performances, such as noble gas and noble liquid primary and secondary scintillations in the NIR, electron emission through the liquid-gas interface in two-phase systems, electron avalanching at low temperatures and noble liquid electroluminescence.

Further studies in the field of Cryogenic Avalanche Detectors are in progress.

## 6. Acknowledgements

I am grateful to D. Akimov, A. Bondar, A. Breskin, A. Chegodaev, A. Grebenuk, E. Shemyakina, R. Snopkov, A. Sokolov and Y. Tikhonov for collaboration in the field of Cryogenic Avalanche Detectors and to A. Bolozdynya, Y. Ramachers, A. Rubbia, J.M.F. dos Santos and D. Thers for useful discussions in this area. I am also grateful to organizers of the MPGD2011 conference, in particular to A. Breskin and A. Ochi, for inviting me to give this review. This work was supported in part by special Federal Program "Scientific and scientific-pedagogical personnel of innovative Russia" in 2009-2013 and by Grant of the Government of Russian Federation (11.G34.31.0047).